\documentclass[conference]{IEEEtran}
\IEEEoverridecommandlockouts
\usepackage{cite}
\usepackage{amsmath,amssymb,amsfonts}
\usepackage{algorithmic}
\usepackage{graphicx}
\usepackage{textcomp}
\usepackage{xcolor}
\usepackage{lastpage}
\usepackage{fancyhdr}

\def\BibTeX{{\rm B\kern-.05em{\sc i\kern-.025em b}\kern-.08em
    T\kern-.1667em\lower.7ex\hbox{E}\kern-.125emX}}
\begin{document}

\title{\textbf{Understanding Public Safety Trends in Calgary through data mining}\\
}

\author{\IEEEauthorblockN{Zack Dewis}
\IEEEauthorblockA{\textit{Dept. of Geomatics} \\
\textit{University of Calgary}\\
Calgary, Alberta \\
zachary.dewis@ucalgary.ca \\
30091912}
\and
\IEEEauthorblockN{Apratim Sen}
\IEEEauthorblockA{\textit{Dept. of Geomatics} \\
\textit{University of Calgary}\\
Calgary, Alberta \\
apratim.sen@ucalgary.ca \\
30202853}
\and
\IEEEauthorblockN{Jeffrey Wong}
\IEEEauthorblockA{\textit{Dept. of Geomatics} \\
\textit{University of Calgary}\\
Calgary, Alberta \\
chunkit.wong@ucalgary.ca \\
30200496}
\and
\IEEEauthorblockN{Yujia Zhang}
\IEEEauthorblockA{\textit{Dept. of Geomatics} \\
\textit{University of Calgary}\\
Calgary, Alberta \\
yujia.zhang2@ucalgary.ca \\
30221253}
}
\maketitle

\begin{abstract}
This paper utilizes statistical data from various open datasets in Calgary to to uncover patterns and insights for community crimes, disorders, and traffic incidents. Community attributes like demographics, housing, and pet registration were collected and analyzed through geospatial visualization and correlation analysis. Strongly correlated features were identified using the chi-square test, and predictive models were built using association rule mining and machine learning algorithms. The findings suggest that crime rates are closely linked to factors such as population density, while pet registration has a smaller impact. This study offers valuable insights for city managers to enhance community safety strategies.
\end{abstract}

\section{Introduction}
\subsection{Research Background and Objectives}
With the acceleration of urbanization,community crimes, disorders, and traffic incidents have become increasingly prominent issues, severely impacting residents' quality of life and the sustainable development of cities. Data mining techniques provide new perspectives for addressing these problems. This study, using Calgary, Canada, as an example, aims to utilize multi-source urban data including streetlights, trees, traffic accidents, crimes, pet registrations, census data, community security, and traffic camera data to explore the key factors affecting community safety and construct predictive models, providing data support for urban management decisions.
Clustering algorithms such as K-Means and DBSCAN are commonly used in urban data mining, but their clustering effectiveness for complex data needs improvement. This paper proposes to introduce advanced clustering algorithms such as CLARANS and CLIQUE, and enhance clustering performance through parameter optimization.
\subsection{Research Questions and Innovations}
This study aims to address the following questions:
(1) What are the characteristics of public facilities, population structure, crime, and accident distribution in various communities in Calgary?
(2) Which community factors are significantly correlated with safety issues?
(3) Can effective predictive models for crime, security incidents, and traffic accidents be constructed?

The innovations of this study include:
(1) Aggregating cutting-edge urban data such as streetlights, trees, and pets to deconstruct the multidimensional factors influencing community safety;
(2) Introducing advanced clustering algorithms such as CLARANS and CLIQUE, integrating optimized clustering with predictive analysis;
(3) Establishing predictive models based on real data, providing new insights for fine-grained urban governance.
\section{Literature Review}
The increasing prevalence of community crimes, disorders, and traffic incidents due to rapid urbanization has become a major concern for city dwellers and urban planners alike. These issues not only affect the quality of life of residents but also pose significant challenges to the sustainable development of cities. In recent years, data mining techniques have emerged as a promising approach to tackle these problems by providing new perspectives and insights. The present study focuses on Calgary, Canada, and aims to identify the key factors influencing community safety by analyzing multi-source urban data and developing predictive models that can support decision-making in urban management \cite{g, i, j}.

K-Means and DBSCAN are among the most widely used clustering algorithms in urban data mining. However, their performance on complex data sets often leaves room for improvement. To address this issue, the current study explores the use of advanced clustering algorithms, such as CLARANS and CLIQUE, and seeks to optimize their performance through parameter tuning. This work builds upon the foundations laid by Ng and Han \cite{c} and Hartigan and Wong \cite{a} in the field of clustering algorithms. Moreover, the efficiency analysis of the K-Means algorithm conducted by Kanungo et al. \cite{b} offers valuable insights into the data-sensitive nature of the algorithm's running time, which is particularly relevant to the data sets used in this study.

When it comes to analyzing crime data, regression analysis is a commonly used approach. The methods employed in this study draw inspiration from the correlated data regression analysis proposed by Liang and Zeger \cite{f}, which is particularly useful when dealing with the inherent correlation issues often encountered in biomedical research data. In addition, Schubert et al.'s \cite{d} critical review of the DBSCAN algorithm, which highlights the importance of selecting appropriate parameters for optimal performance, is taken into consideration. The density-based clustering structure of data sets, as demonstrated by Ankerst et al. \cite{e} using the OPTICS algorithm, also serves as a guiding principle for the approach adopted in this study.

The methodology of this study is further informed by the work of other researchers in the field of spatial data mining. For instance, Bernasco and Elffers \cite{k} provide a comprehensive overview of the statistical analysis of spatial crime data, while Anselin et al. \cite{l} demonstrate the application of spatial analysis in crime research. These studies not only introduce new methods for analyzing spatial data but also shed light on the intricate relationship between crime and place.

In the era of big data, data integration presents a significant challenge that must be addressed. Dong and Srivastava \cite{g} delve into the progress made by the data integration community in tackling issues such as schema mapping, record linkage, and data fusion when working with large, dynamic, heterogeneous, and varying quality data sources. By incorporating cutting-edge urban data, such as information on streetlights, trees, and pets, this study aims to contribute to the understanding of the multifaceted factors that influence community safety in Calgary.

Although the primary focus of this study is on modeling large-scale, global structures, the principles and techniques discussed by Hand \cite{m} in the context of detecting adverse drug reactions in the pharmaceutical industry can also be applied to identify anomalies and outliers in urban crime data. This approach contributes to a more comprehensive understanding of community safety.

While this study makes notable contributions to the field, it is not without limitations. Due to constraints in data availability, the study only covers a limited set of community indicators and spans a relatively short time frame. Future research could benefit from collecting more comprehensive data and conducting long-term spatiotemporal analyses. Additionally, the complex and diverse nature of the causes underlying urban safety issues makes it challenging to establish clear causal relationships and mechanisms based solely on data analysis. To address this, subsequent studies could incorporate qualitative methods, such as expert interviews and case studies, to explore the socioeconomic, institutional, and cultural factors that influence community safety.

This study builds upon existing algorithms and methods from the literature, making improvements and extensions to enable more effective processing and analysis of urban safety data. By combining various clustering and regression analysis techniques into predictive models, this research provides valuable scientific evidence and insights that can inform urban governance and contribute to the development of safer cities. The findings underscore the crucial role that data mining techniques can play in addressing the challenges posed by urbanization and promoting the well-being of city residents.

\section{Data Source and Preprocessing}
The data used in this study were obtained from the City of Calgary's open data platform, including 2019 data on streetlights, public trees, traffic incidents, crime statistics, licensed pets, census by community, community disorder statistics, and traffic cameras.

\subsection{Data Cleaning and Geocoding}
Despite all of the data being acquired from the same source, data cleaning was still necessary in order to extract interesting patterns from the data and to perform further analysis. Starting off, the data was manually inspected, where any columns that were deemed unnecessary or blank were removed. Next, all the datasets were loading into Panda dataframes for easy further cleaning. At this stage all rows that were duplicate entries to another row in any of the seven datasets were removed. For the columns that had either Na values of missing values, these were modified on a case-by-case basis. If the column has located in the crime dataset, the values were replaced with zero as the assumption was made that this meant there was zero crime in this particular neighbourhood. This occurred for the neighbourhoods that were still being developed in Calgary when the 2019 census was being conducted and therefore any crime that may have occurred was deemed irrelevant to the research goal.  

Another issue that arose during the cleaning process was that not all the datasets provided latitude and longitude coordinates for where each of the attributes was located. This was solved by running a python script to join the datasets and assign each a latitude and longitude to allow for clustering analysis. Some datasets contained both latitude and longitude as well as a community name or identifier, some contained just a community name and some only contained latitude and longitude. The decision was made that each dataset should be given both latitude and longitude coordinates and a community name attached to it. To achieve this, three different methodologies were used depending on what each individual dataset already contained. If the dataset already contained latitude, longitude and a community name, no further action was required. For datasets only containing a community name, the community centroid coordinates were assigned to the data to provide it a latitude and longitude. For datasets only containing latitude and longitude, a community boundary shapefile was used, where a script was then ran to see which community the points intersected and then a name was assigned. 

At this point all of the necessary data cleaning had been performed for analysis to be carried out. Geocoding had also been performed on all of the datasets, allowing for summary statistics to then be calculated to perform correlation and regression analysis. 

\subsection{Feature Engineering and Data Integration}

With the data cleaning and geocoding complete, the next step was to aggregate the data to the same level in order to perform analysis. Community level was selected as the resolution for all the data to be aggregated to. The reason for this was that it was the finest resolution that was available to be used without having to make assumptions and convert datasets to finer resolutions than they are provided at. As the community census and crime was only at community level, a more fine solution could not be used for aggregation. 

At the community level, we calculated various indicators such as the number of streetlights, average wattage, total wattage, number of trees, number of traffic incidents, total number of crimes and crime categories, number of community disorder events, total number of registered pets, and number of registered cats and dogs. We also extracted demographic features of communities from census data, such as total population, male-female ratio, number of dwellings, and number of apartments.

Finally, all of these statistical indicators were compiled into a new geopanda dataframe to be used for further analysis. The original dataframes were also maintained and used for cluster analysis as many were at a finer resolution and allowed for clustering to be performed on them. The data is now fully ready for data mining processes to be performed with the goal of being able to extract interesting patterns and insight from it. 

\section{Methodology}

To best extract patterns from the dataset, several exploratory data analyses and different data mining techniques were used in hopes of extracting interesting patterns. This includes data summarization, visualization, temporal analyses, clustering, correlation, regression, and association rule mining. 

\subsection{Exploratory Data Analysis}
Exploratory Data Analysis (EDA) is a crucial step in data analysis, especially when dealing with new datasets. It involves exploring and summarizing the main characteristics, patterns, and relationships present in the data. The primary goal of EDA is to gain insights and understanding of the data to inform further analysis or modeling tasks.
\subsubsection{Data Summarization}
Firstly, we started with summarizing the main characteristics of the data, by categories and by community, for the total number of categorical features. For the traffic incident dataset, there were no categorical data but only the descriptions. We mapped some keywords in the description column to a new column 'category'.
\subsubsection{Geospatial Visualization}
And visualizations of the data were created and manually analyzed to see if there were any regions of interest within the city that should have further inspection or analysis performed. To visually demonstrate the distribution differences of various indicators among communities, we created interactive maps using the Folium library. Specifically, we used community polygons as basic plotting units, colored the polygons based on the values of each indicator, and added legend explanations. Additionally, we experimented with various color schemes to better highlight spatial patterns.
\subsubsection{Temporal Analysis}
Temporal analysis were performed to examine how community crimes, disorders, and traffic incidents change over time. 

\subsection{Correlation Analysis}
Then, correlation mining was also performed to determine which attributes were positively or negatively correlated with each other within the city. 
\subsubsection{Correlation Analysis}
To explore the correlations between factors and safety issues, we computed the correlation matrix of gdf\_comm and plotted heatmaps using the Seaborn library. We focused on the correlations between demographic indicators, indicators related to children and schools, streetlight and tree indicators, pet quantity indicators, and crimes, security incidents, and traffic accidents

\subsection{Cluster Analysis}

To further explore the intrinsic structure and similarities among communities, we employed various clustering algorithms onto the cleaned geocoded datasets. In selecting clustering algorithms the goal was to cover as many different approaches as possible and see what would have the best results on the data. Based on our research \cite{Clustering}, we found that density-based and grid-based methods are usually more suitable for handling spatial data. For the partitioning approach, K-Means and CLARANS were selected, Agglomerate Clustering was chosen for the hierarchical approach, DBSCAN and OPTICS were selected for the density based approach and CLIQUE was used for the grid-based approach. Overall, the approach for how each method was implemented is quite similar, but nonetheless will be broken down into more detail below. 

\subsubsection{K-Means Clustering}
K-Means was the first algorithm implemented. It was implemented on each dataset using the latitude and longitude coordinates to determine clusters. To determine the optimal number of clusters, we used the silhouette score as the evaluation metric, tested a range of 2 to 20 clusters, and plotted the relationship between the silhouette score and the number of clusters. Finally, we selected the optimal value corresponding to the highest silhouette score. The clustering results were visualized using Folium maps.
\newline
\begin{figure}
    \centering
    \fbox{\begin{minipage}{23em}
    \textbf{Step 1:} Accept the number of clusters to group data into and the dataset to cluster as input values.\\
    \textbf{Step 2:} Initialize the first K clusters.\\
    \textbf{Step 2.1:} Take the first k instances or.\\
    \textbf{Step 2.2:} Take a random sampling of k elements.\\
    \textbf{Step 3:} Calculate the arithmetic means of each cluster formed in the dataset.\\
    \textbf{Step 4:} K-means assigns each record in the dataset to only one of the initial clusters.\\
    \textbf{Step 4.1:} Each record is assigned to the nearest cluster using a measure of distance (e.g., Euclidean distance).\\
    \textbf{Step 5:} K-means re-assigns each record in the dataset to the most similar cluster and re-calculates the arithmetic mean of all the clusters in the dataset. \cite{Kmeans}
    \end{minipage}}
    \caption{Generalized pseudocode of traditional K-means}
    \label{fig:kmeans_algorithm}
\end{figure}

\subsubsection{CLARANS Clustering}
Considering that the K-Means algorithm is susceptible to outliers, we further attempted the CLARANS (Clustering Large Applications based upon RANdomized Search) algorithm. This algorithm minimizes the cost function through random search and can efficiently handle large datasets. We implemented CLARANS clustering using the pyclustering library and conducted grid search on different parameter combinations to achieve the optimal clustering effect. Again, the silhouette score was used to determine what the optimal parameters for the algorithm were. This was repeated for each dataset to determine the optimal parameters. 
\newline
\begin{figure}
    \centering
    \fbox{
    \begin{minipage}{23em}
        \textbf{Step 1:} Input parameters numlocal and maxneighbour.\\
        \textbf{Step 2:} Set current to an arbitrary node in $G(n,k)$.\\
        \textbf{Step 3:} Set $j$ to 1.\\
        \textbf{Step 4:} Consider a random neighbor $S$ of current, calculate the cost differential of the two nodes.\\
        \textbf{Step 5:} If $S$ has a lower cost, set current to $S$, and go to Step 3.\\
        \textbf{Step 6:} Otherwise, increment $j$ by 1. If $j \geq$ maxneighbour, go to Step 4.\\
        \textbf{Step 7:} Otherwise, when $j >$ maxneighbour, compare the cost of current with mincost. If the former is less than mincost, set mincost to the cost of current, and set bestnode to current.\\
        \textbf{Step 8:} Increment $i$ by 1. If $i >$ numlocal, output bestnode and halt. Otherwise, go to Step 2. \cite{CLARANS}
    \end{minipage}}
    \caption{Generalized version of CLARANS algorithm}
    \label{fig:clarans_algorithm}
\end{figure}

\subsubsection{CLIQUE Clustering}
To explore hidden patterns in high-dimensional data, we also introduced the grid-based density clustering algorithm CLIQUE (Clustering In QUEst). This algorithm first divides the data space into a series of grid cells, then identifies high-density cells, and finally forms clusters by combining adjacent high-density cells. We implemented CLIQUE clustering using the pyclustering library and conducted sensitivity analysis on parameters such as cell size and density threshold.
\subsubsection{DBSCAN Clustering}
To assess a density based clustering method, DBSCAN was implemented on each of the datasets, again with the silhouette score being used to determine the optimal parameters DBSCAN functions by categorizing points as either core, border or outlier/noise points. This is determined based on the density of the points, a point becomes a core point if it has at least the minimum amount of points needed to reach the threshold set by the user within its boundary (also set by the user). Multiple different parameter sets were selected, for both the threshold and the minimum amount of points needed for a point to become a core point.
To explain the algorithm, we first need to define density connected point: Two points are called density connected if there is a core point which is density reachable (i.e. connected through a series of core points) from both the points.
\begin{figure}
    \centering
    \fbox{\begin{minipage}{23em}
    \textbf{Step 1:} Choose a value for eps and MinPts.\\
    \textbf{Step 2:} For a particular data point (x), calculate its distance from every other data point.\\
    \textbf{Step 3:} Find all the neighborhood points of x which fall inside the circle of radius (eps) or simply whose distance from x is smaller than or equal to eps.\\
    \textbf{Step 4:} Treat x as visited and if the number of neighborhood points around x is greater or equal to MinPts, then treat x as a core point. If it is not assigned to any cluster, create a new cluster and assign it to that.\\
    \textbf{Step 5:} If the number of neighborhood points around x is less than MinPts and it has a core point in its neighborhood, treat it as a border point.\\
    \textbf{Step 6:} Include all the density-connected points as a single cluster. (What density-connected points mean is described later).\\
    \textbf{Step 7:} Repeat the above steps for every unvisited point in the dataset and find out all core, border, and outlier points. \cite{DBSCAN}
    \end{minipage}}
    \caption{Generalized version of DBSCAN algorithm}
    \label{fig:dbscan_algorithm}
\end{figure}

\subsubsection{OPTICS Clustering}
Optics clustering is also a density based clustering algorithm, but does not require a minimum amount of points or a distance threshold to function. Instead it requires a reach-ability distance. This creates a reach-ability plot to determine what points can be reached from each other. The results of this are another way to showcase how dense the distribution of points in the dataset are. 
\begin{figure}
    \centering
    \fbox{
    \begin{minipage}{23em}
        \textbf{Step 1:} Mark every point of the dataset as 'unclustered'. Corresponding to each point, mark both core distances and reachability distances as 'undefined'. Clusters = 1.\\
        \textbf{Step 2:} Pick any point $P \in X$ (the dataset) and evaluate $N_\epsilon(P)$ ($\epsilon$ Neighborhood of P).\\
        \textbf{Step 2.1:} Compare core distance of P only if $|N_\epsilon(P)| > \text{MinPts}$.\\
        \textbf{Step 2.2:} Assign values of cluster to P once core distance is computed. Compute reachability distance of every point in $N_\epsilon(P)$.\\
        \textbf{Step 2.2.1:} If they are undefined, replace undefined with computed reachability distance.\\
        \textbf{Step 2.2.2:} If a reachability distance exists for that point, replace it with the newly computed reachability distance only if new reachability distance $<$ old reachability distance.\\
        \textbf{Step 3:} We consider every unclustered element in $N_\epsilon(P)$ and repeat Step 2.\\
        \textbf{Step 4:} When all elements of $N_\epsilon(P)$ are clustered, along with the neighborhood of each point of $N_\epsilon(P)$ [Step 3], we do the following:\\
        \textbf{Step 4.1:} $P$ = unclustered point of $X$.\\
        \textbf{Step 4.2:} Cluster = Cluster + 1.\\
        \textbf{Step 4.3:} Repeat Step 2 to Step 3. We will keep on repeating the steps until the entire dataset is clustered.\cite{Optics}
    \end{minipage}}
    \caption{Generalized version of OPTICS algorithm}
    \label{fig:optics_algorithm}
\end{figure}

\subsubsection{Hierarchical Clustering}
Hierarchical clustering builds a hierarchy of clusters, where the agglomerative clustering method was chosen. This functions by selecting a single data point as a single cluster and then merges pairs of clusters until only one remains. Agglomerative clustering uses a tree-like structure called a dendrogram, which is then cut to determine the optimal number of clusters at different levels. The advantage to hierarchical clustering is that it does not require the number of clusters to be specified in advance, where the data does not need to naturally fall into a specific number of clusters to produce a high quality result. 
\begin{figure}
    \centering
    \fbox{
    \begin{minipage}{23em}
        \textbf{Step 1:} Each data point is its own cluster. For N data points, we have N clusters.\\
        \textbf{Step 2:} Find the inter-cluster distance between every pair of clusters.\\
        \textbf{Step 3:} Join the two clusters having the lowest inter-cluster distance.\\
        \textbf{Step 4:} Repeat Step 2 until either:\\
        \textbf{Step 4.1:} A set point or threshold is reached.\\
        \textbf{Step 4.2:} No more clustering can be done.
    \end{minipage}}
    \caption{Generalized version of Agglomerative Hierarchical Clustering algorithm}
    \label{fig:agglomerative_algorithm}
\end{figure}

\subsection{Predictive Analysis}
To predict the safety risks of communities, we constructed multiple linear regression and random forest regression models.
\subsubsection{Data Splitting and Feature Selection}
We first used the chi-square test to select variables significantly correlated with the dependent variables (crime, security incidents, traffic accidents). Then, we randomly split the data into training set (80\%) and testing set (20\%) for model training and evaluation.
\subsubsection{Multiple Linear Regression}
We constructed multiple linear regression models using the Sklearn library, with the number of crimes, number of security incidents, and number of traffic accidents as dependent variables and the selected significantly correlated indicators as independent variables. We fitted the models on the training set and evaluated them on the testing set using mean squared error and coefficient of determination.
\subsubsection{Random Forest Regression}
Considering the possible nonlinear relationships between variables, we further attempted random forest regression models. Random forest combines multiple decision trees to effectively capture complex interactions between variables. We implemented random forest regression using the Sklearn library and optimized hyperparameters such as the number of trees and maximum depth through grid search. We also calculated the feature importance of each variable to assess its impact on prediction results.
\subsubsection{Model Evaluation and Comparison}
We evaluated the predictive performance of multiple linear regression and random forest regression models on the testing set, using mean squared error and coefficient of determination as evaluation metrics. We also plotted scatter plots of actual values versus predicted values to visually demonstrate the model fitting effects. Finally, we conducted comparative analysis of the performance of the two types of models.

\section{Results}
\subsection{Data Summarization}
Fig. 6 reveals a varied landscape of crime in Calgary, with theft from vehicles dominating as the most prevalent offense, followed by theft of vehicles and break and enter incidents involving commercial properties. Assault, both domestic and non-domestic, also remains a concerning issue, with a notable number of reported incidents. Robbery, though relatively lower in count, still poses a risk in public and commercial settings. 
\begin{figure}
    \centering
    \includegraphics[width=1\hsize]{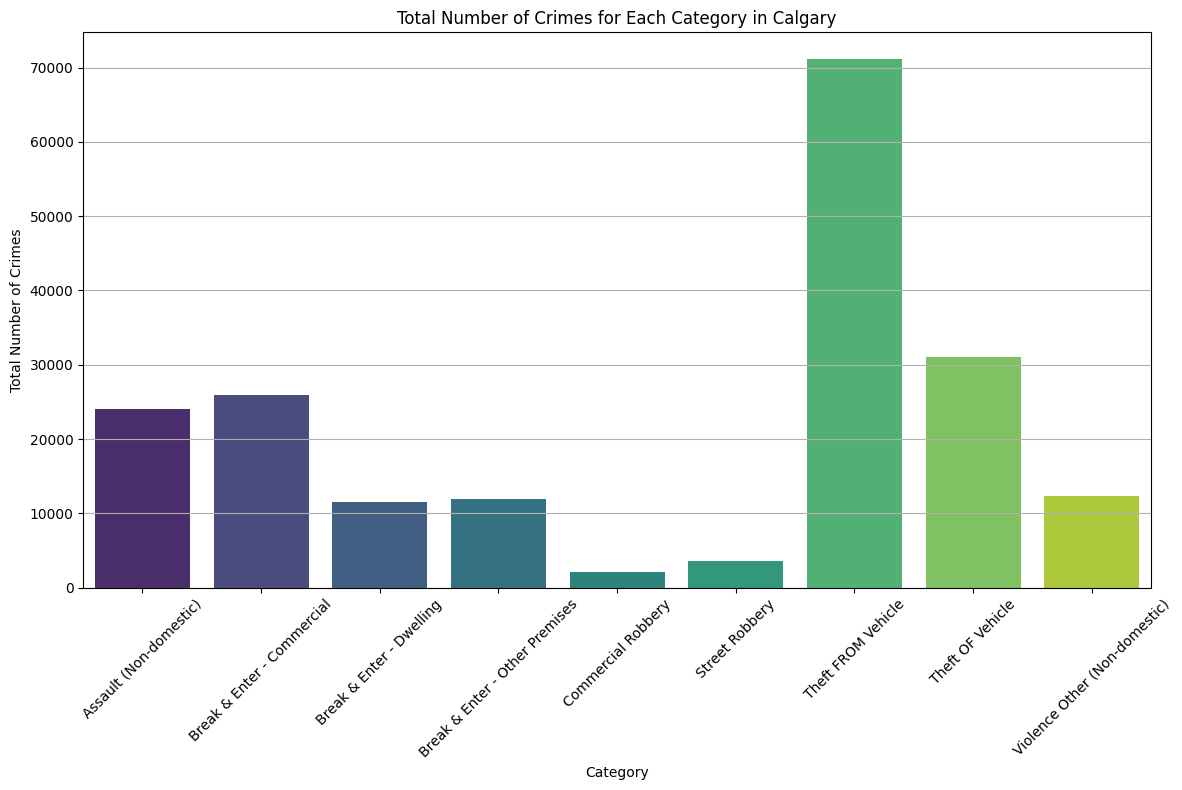}
    \caption{Total Number of Crimes for Each Category in Calgary}
    \label{fig:total_crimes}
\end{figure}

Fig. 7 reveals the top 10 traffic incident categories in Calgary, with "2 vehicle incident" and "Traffic Incident" leading. Issues like road blocking, multi-vehicle incidents, and pedestrian involvement are also significant. 
\begin{figure}
    \centering
    \includegraphics[width=1\hsize]{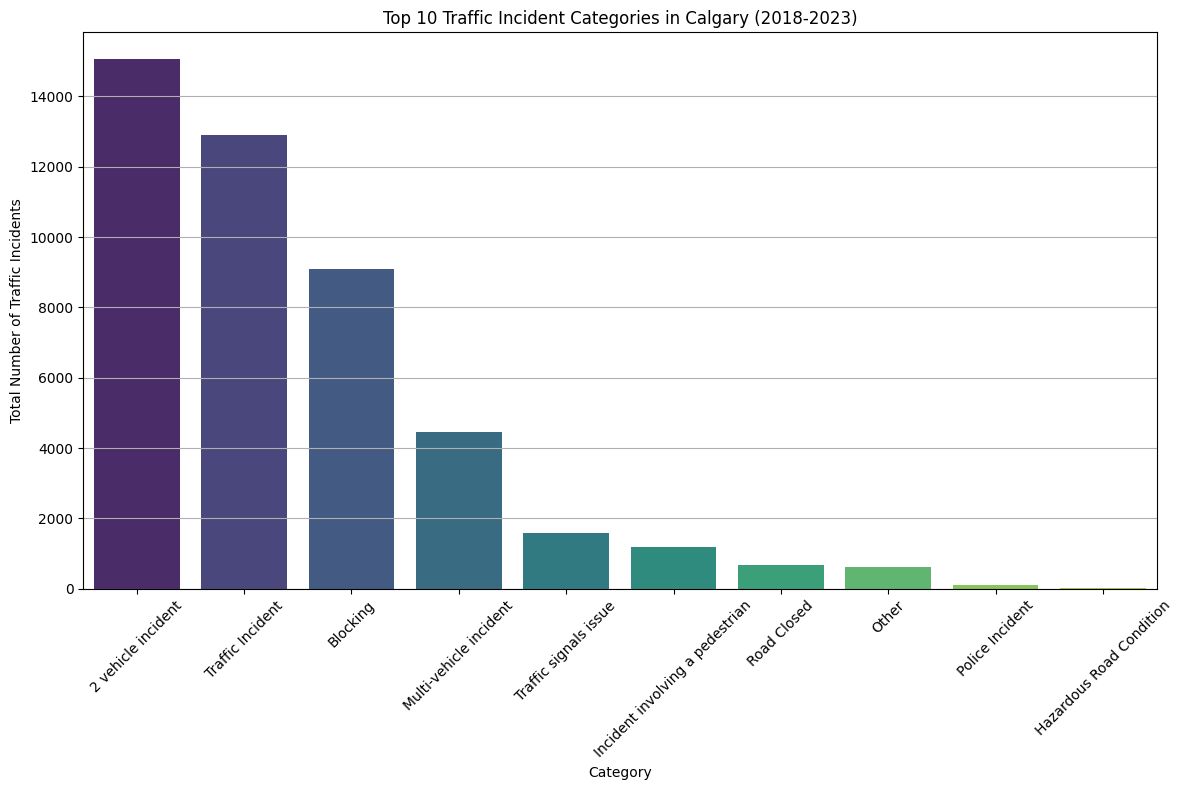}
    \caption{Top 10 Traffic Incident Categories in Calgary (2018-2023)}
    \label{fig:total_traffic}
\end{figure}

In Fig 8 and 9, we can easily see that Beltline and Downtown are the two leading communities with highest total number of crimes and disorders in Calgary. And the Forest Lawn community is also notable that comes next.
\begin{figure}
    \centering
    \includegraphics[width=1\hsize]{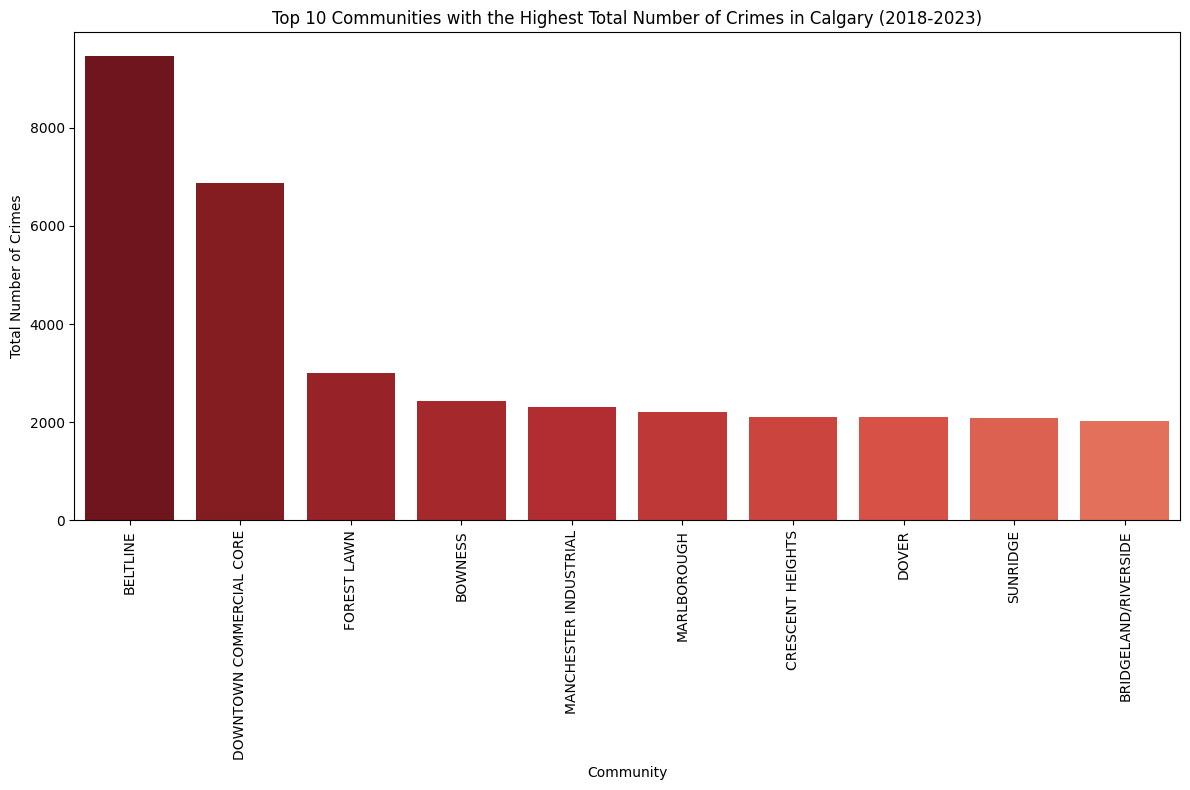}
    \caption{Top 10 Communities with the Highest Total Number of Crimes in Calgary (2018-2023)}
    \label{fig:top10_crime}
\end{figure}

\begin{figure}
    \centering
    \includegraphics[width=1\hsize]{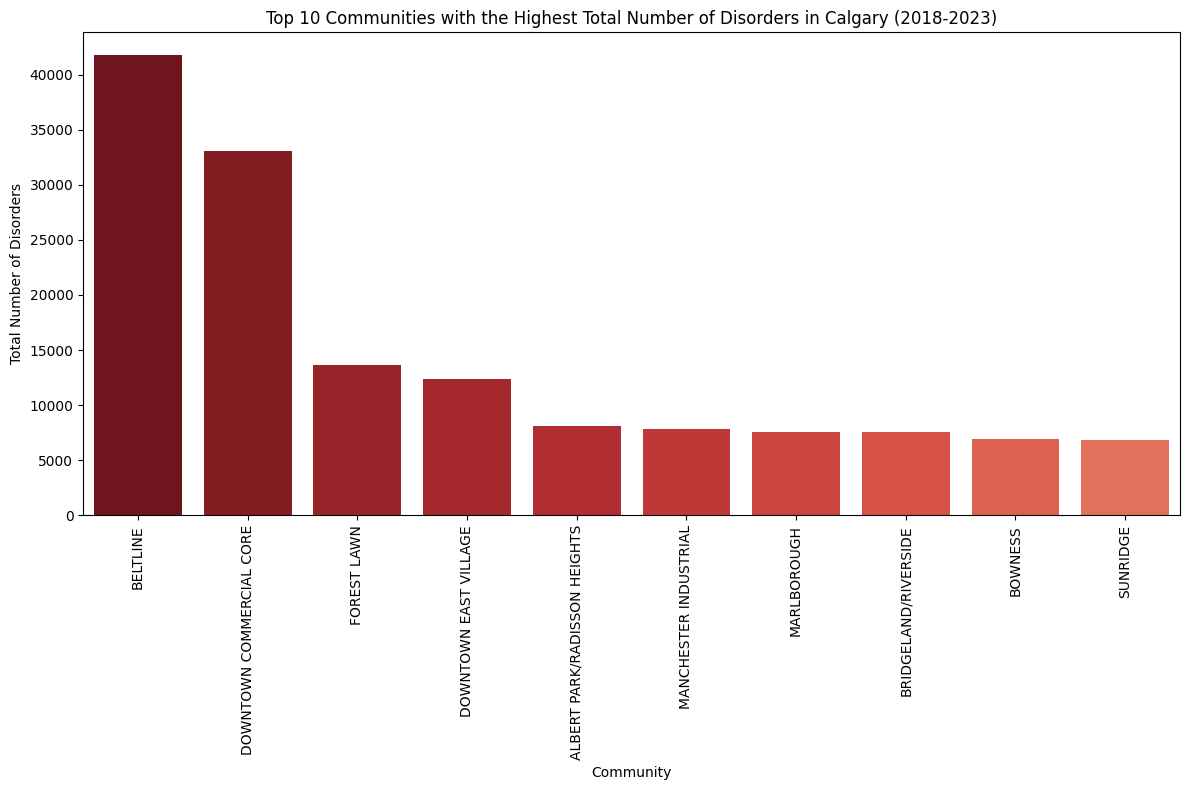}
    \caption{Top 10 Communities with the Highest Total Number of Disorders in Calgary (2018-2023)}
    \label{fig:top10_disorder}
\end{figure}

Fig. 10 highlights the top 10 communities in Calgary with the highest number of traffic incidents. Bridgeland/Riverside leads with 1,244 incidents, followed closely by Beltline and Downtown Core with 1,153 and 1,103 incidents respectively. Burns Industrial, Albert Park, and Manchester Industrial also rank among the communities with notable traffic incident counts.
\begin{figure}[ht]
    \centering
    \includegraphics[width=1\hsize]{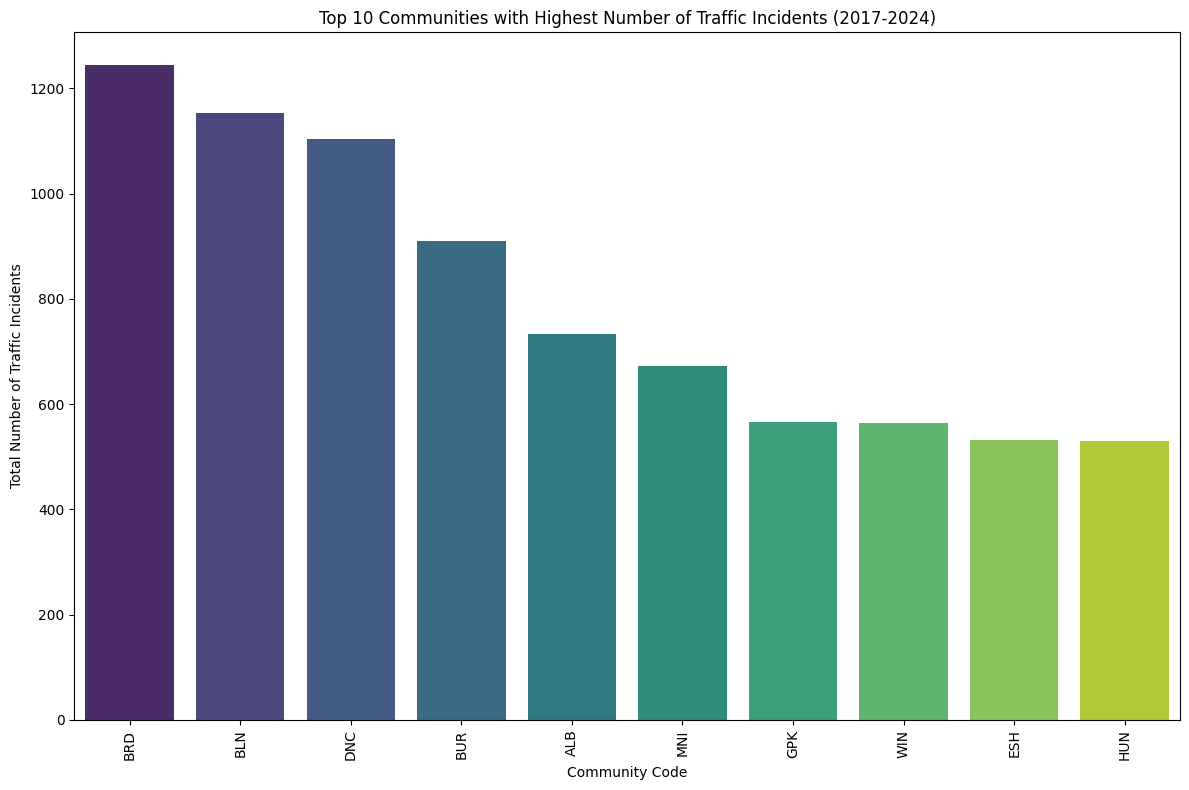}
    \caption{Top 10 Communities with Highest Number of Traffic Incidents (2017-2024)}
    \label{fig:top10_traffic}
\end{figure}

\subsection{Map Visualization}
Fig. 11 shows the map visualization with multi-polygons colored by the number of overall crimes in Calgary in each community. From a geospatial point of view, we can see that large amount of Crimes happened in the centre areas of Calgary, which are the Beltline and Downtown Core. We can also see in the visualization that the communities in the Northeast sector are with slightly more crimes than communities in other sectors.
\begin{figure}
    \centering
    \includegraphics[width=1\hsize]{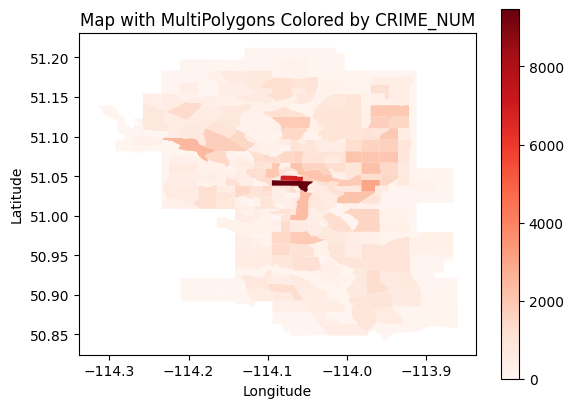}
    \caption{Map with MultiPolygons Colored by number of Crimes in Calgary}
    \label{fig:crime_map}
\end{figure}

Fig. 12 shows the map visualization with multi-polygons colored by the number of overall disorders in Calgary in each community. Similar to fig. 11, we can see that large amount of Crimes happened in the centre areas of Calgary, which are the same Beltline and Downtown Core. And for the Northeast sector, the number of disorders are also slights higher than other sectors but not as obvious as previous map visualization for crimes.
\begin{figure}
    \centering
    \includegraphics[width=1\hsize]{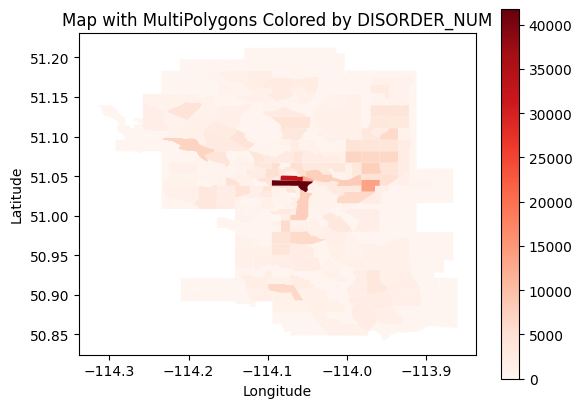}
    \caption{Map with MultiPolygons Colored by number of Disorders in Calgary}
    \label{fig:disorder_map}
\end{figure}

Fig. 13 shows the map visualization with multi-polygons colored by the number of overall traffic incidents in Calgary in each community. Comparing to crimes and disorders, the traffic incidents also have a central tendency in Beltline, Downtown Core with the additional Bridgeland-Riverside community. But the distribution of traffic incidents is more dispersed around the centre areas of Calgary. It is interesting that there seems to be more traffic incidents along the Deerfoot Trail and Macleod Trail.
\begin{figure}
    \centering
    \includegraphics[width=1\hsize]{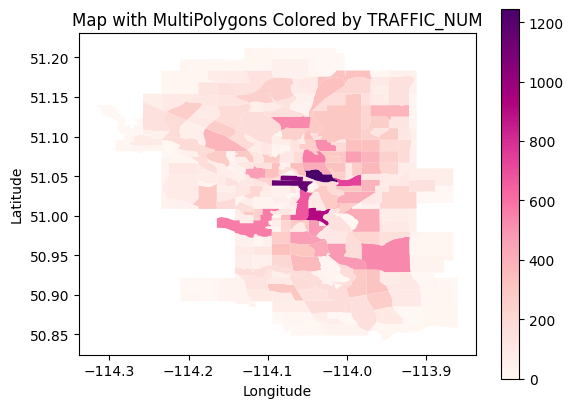}
    \caption{Map with MultiPolygons Colored by number of Traffic Incidents in Calgary}
    \label{fig:traffic_map}
\end{figure}

\subsection{Temporal Analysis}

Fig. 14 shows the Monthly Crime Trend Over Time by Category in Calgary. We can see a slightly decreasing trend on the overall crimes and theft from vehicle from year 2018 towards 2024.
\begin{figure}
    \centering
    \includegraphics[width=1\hsize]{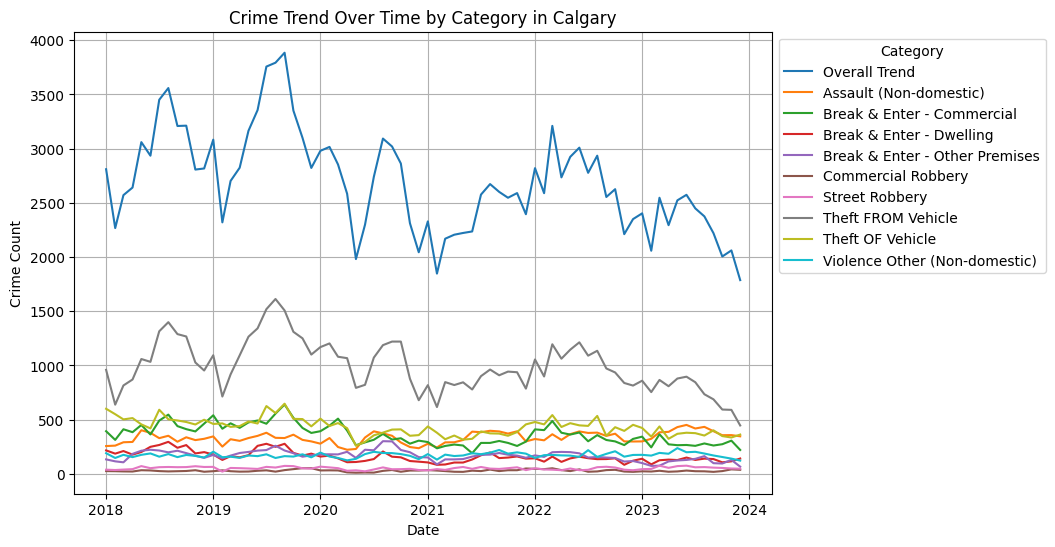}
    \caption{Crime Trend Over Time by Category in Calgary}
    \label{fig:crime_trend}
\end{figure}

While in fig. 15, it is showing Disorder Trend Over Time in Calgary. Instead of a decreasing trend, the disorder trend shows a periodic pattern every year from year 2018 to 2024.
\begin{figure}
    \centering
    \includegraphics[width=1\hsize]{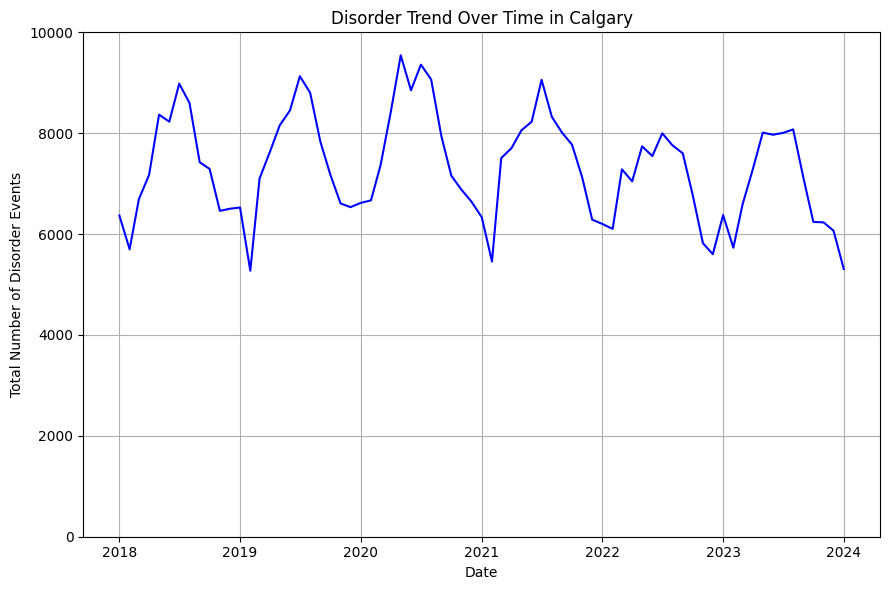}
    \caption{Disorder Trend Over Time in Calgary}
    \label{fig:disorder_trend}
\end{figure}

And in fig. 16, it shows the Traffic Incident Trend Over Time in Calgary. It is observable that there is a slightly increase trend for the number of traffic incidents from year 2017 to 2024.
\begin{figure}
    \centering
    \includegraphics[width=1\hsize]{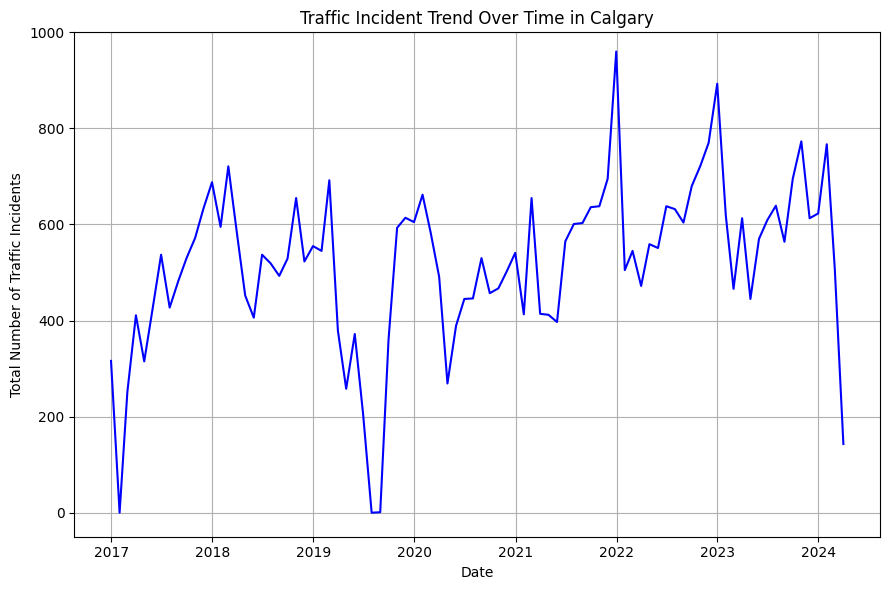}
    \caption{Traffic Incident Trend Over Time in Calgary}
    \label{fig:traffic_trend}
\end{figure}

For fig 17, 18 and 19, they are showing the average Crimes, disorders or traffic incidents per month in Calgary. We can see in fig. 17 that the average number of crimes increases after April with the warmer weather and reach average maximum in August. After that the average number of crimes decreases with the temperature towards December.
\begin{figure}
    \centering
    \includegraphics[width=1\hsize]{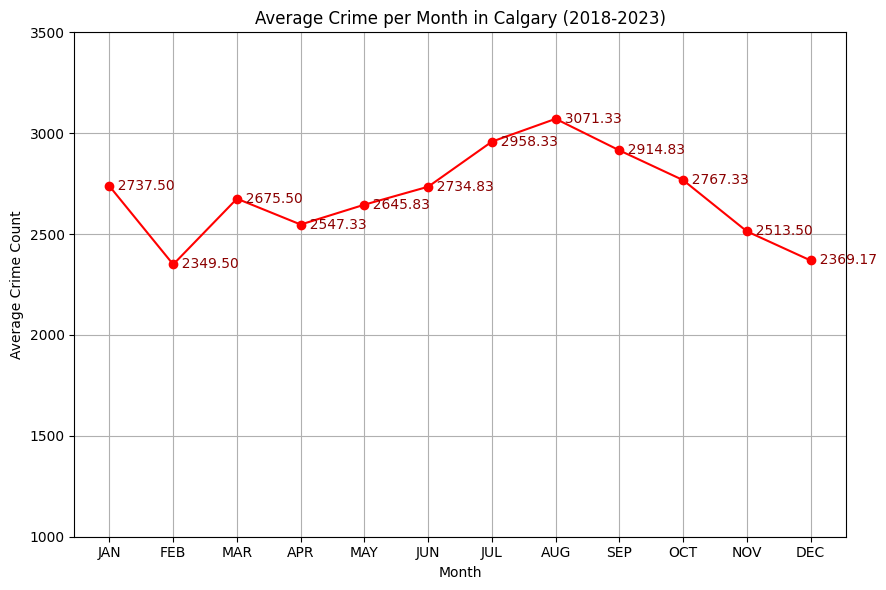}
    \caption{Average Crime per Month in Calgary (2018-2023)}
    \label{fig:crime_monthly}
\end{figure}

While in fig. 18, we can also observe an similar rising trend for the average number of disorders, reaching the maximum in July.
\begin{figure}
    \centering
    \includegraphics[width=1\hsize]{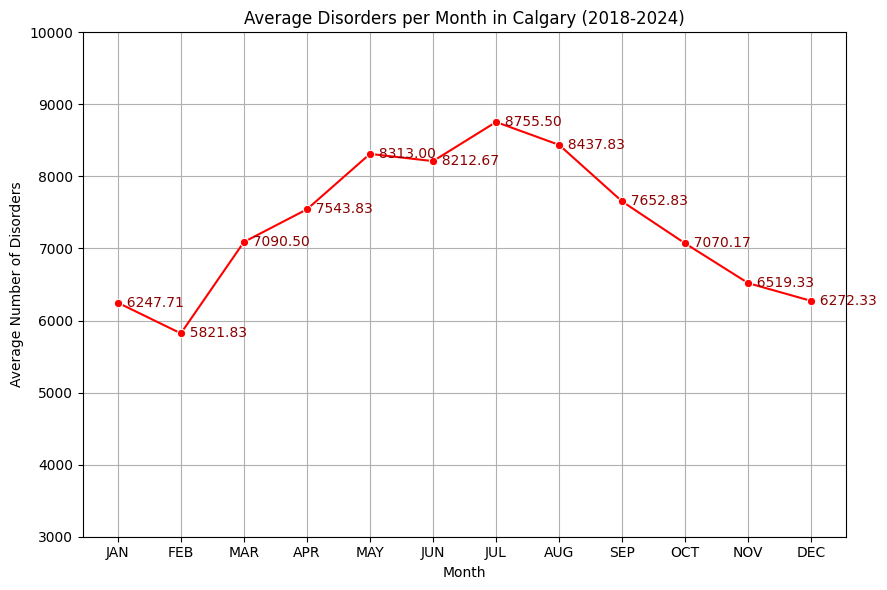}
    \caption{Average Disorders per Month in Calgary (2018-2024)}
    \label{fig:disorder_monthly}
\end{figure}

And for fig. 19, we see a different trend in traffic incidents per month in Calgary. It exhibits an increase trend starting in July towards December which seems to be inversely proportion to the decreasing temperature.
\begin{figure}
    \centering
    \includegraphics[width=1\hsize]{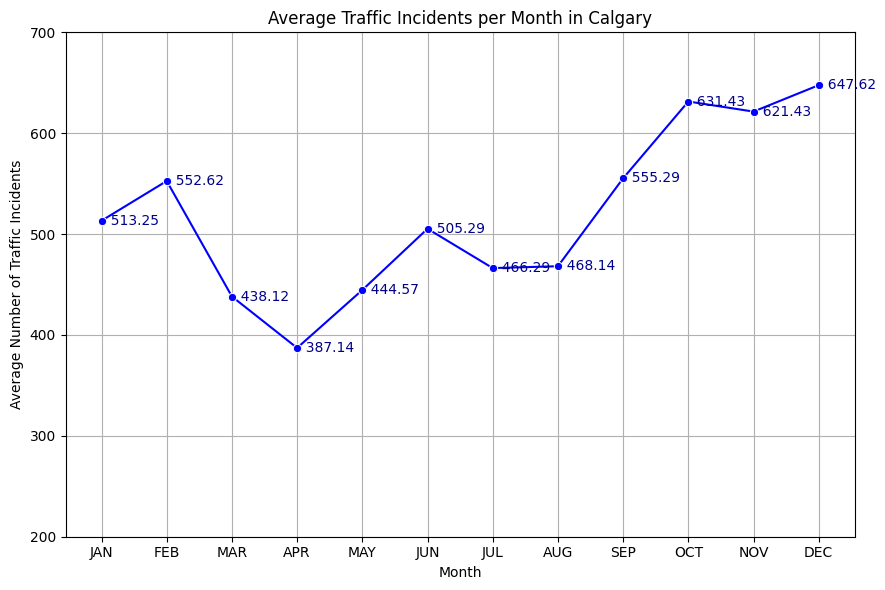}
    \caption{Average Traffic Incidents per Month in Calgary}
    \label{fig:traffic_monthly}
\end{figure}

Fig. 20 shows the Number of Crimes Each Year by Category in Calgary. We can see in this figure that, despite the slight differences in number each year, the pattern on the distribution of crime categories are the same.
\begin{figure}
    \centering
    \includegraphics[width=1\hsize]{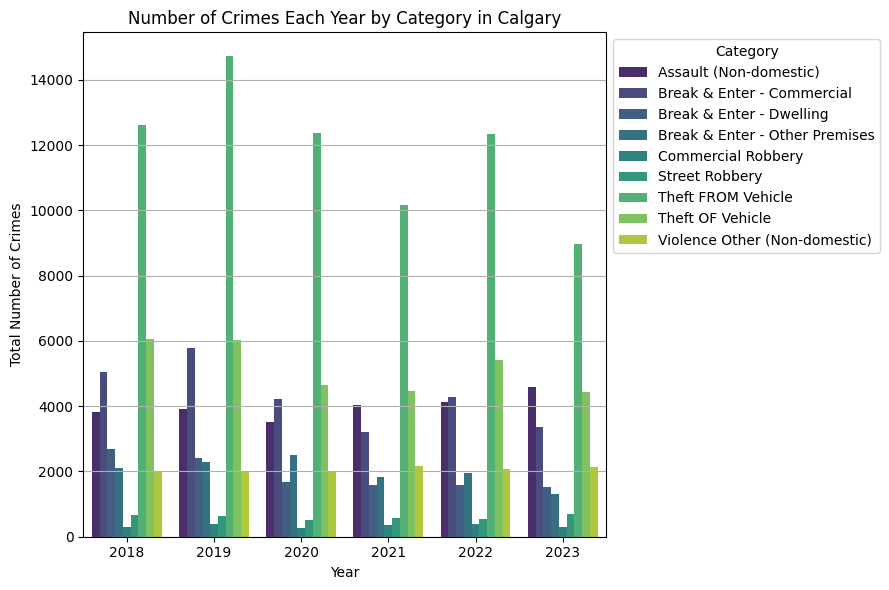}
    \caption{Number of Crimes Each Year by Category in Calgary}
    \label{fig:crime_year_cat}
\end{figure}

Fig. 21 illustrates the total number of disorders in Calgary from 2018 to 2024. Apart from year 2024 with incomplete data, disorder counts remained relatively stable from 2018 to 2023, ranging from 83,701 to 94,514.
\begin{figure}
    \centering
    \includegraphics[width=1\hsize]{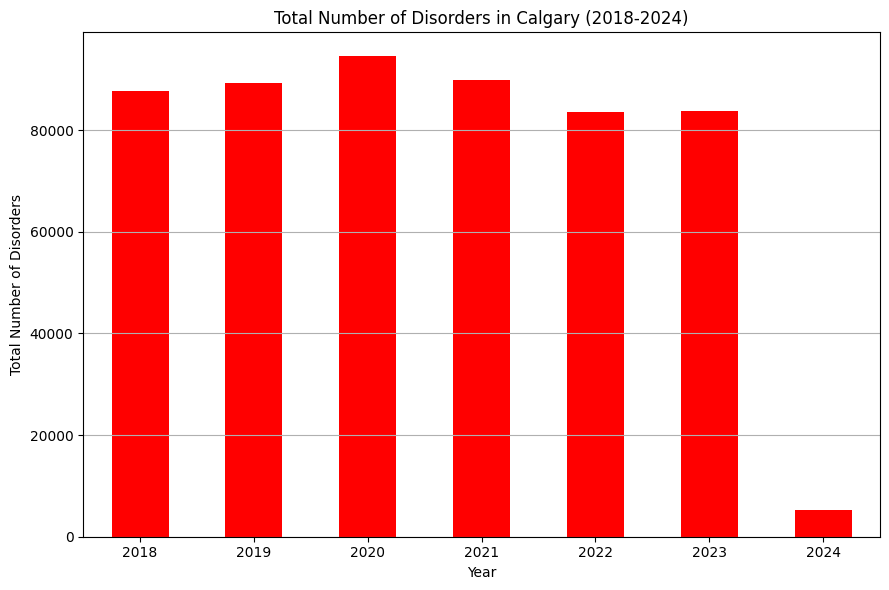}
    \caption{Total Number of Disorders in Calgary (2018-2024)}
    \label{fig:disorder_year}
\end{figure}

Fig 22 shows the top 5 Traffic Incident Categories along the years in Calgary. We can see in the figure that the number of 2 vehicle incidents are decreasing significantly while the blocking and general traffic incidents are increasing. As the category column is only done by mapping key words or phrases from the description column. The trends maybe a result of changing habits in writing more detailed descriptions.
\begin{figure}
    \centering
    \includegraphics[width=1\hsize]{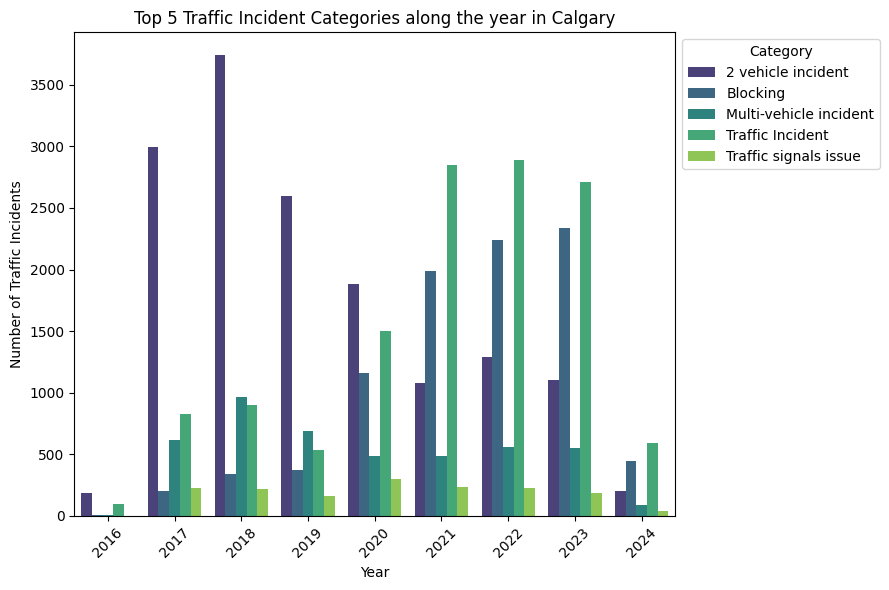}
    \caption{Top 5 Traffic Incident Categories along the year in Calgary}
    \label{fig:traffic_year_cat}
\end{figure}
\subsection{Correlation Analysis Results}
For the correlation analysis, as there are too many attributes in the dataframe of community census, we need to use trial and error method and by our knowledge to screen out attributes that could be related to crimes, disorders and traffic incidents. And we have also tried to screen out attributes that varying across different communities.

Fig 23 is the results of first trial in correlation analysis. It is a correlation heatmap visualization showing the correlations between some attributes in the community census. Despite of availability of correlation results between different attributes in communities, they are not the interest of this project. The interest of this project lies in the last 3 rows which are the correlation of crimes, disorders and traffic incidents in related to community attributes. We can see in the results that the number of crimes and disorders are strongly correlated to the count of dwellings in a community, and it is a much stronger correlation comparing to that of resident counts in a community. Additionally, we can also see that crimes and disorders are strongly correlated to the number of apartments and other apartment related attributes. It indicates that we could probably anticipate more crimes and disorders in a community if there are more apartments in a community or more residents living in an apartment. Between crimes and disorders, they are almost perfectly correlated to each other. While for traffic incidents, there are no obvious correlations to any of the community attributes and it is expected. It is interesting that the number of traffic incidents are relatively correlated to the number of crimes and disorders in a community.
\begin{figure}
    \centering
    \includegraphics[width=1\hsize]{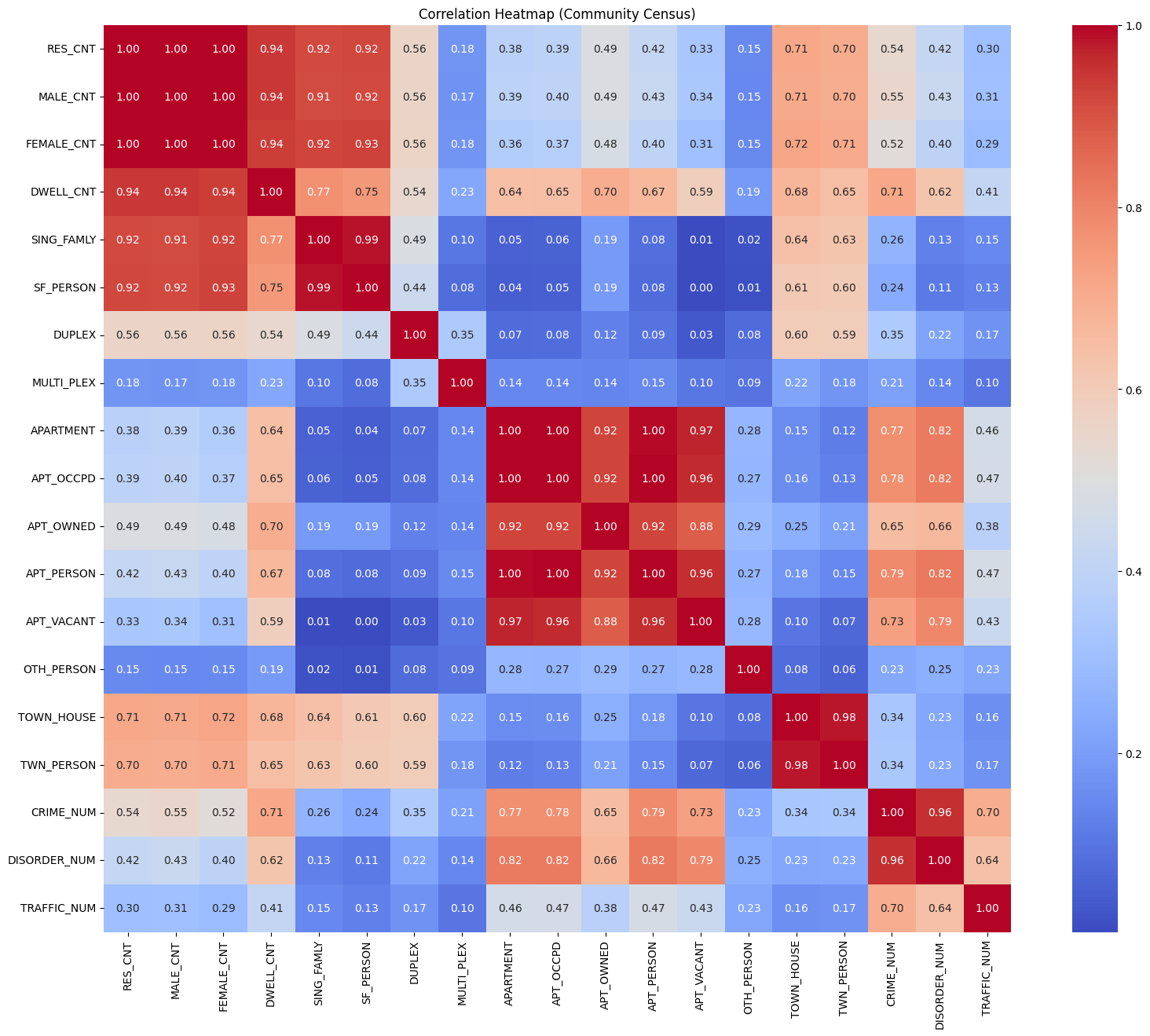}
    \caption{Correlation Heatmap (Community Census)}
    \label{fig:corr_comm_census}
\end{figure}

For fig. 24, we were trying to look into the how the number of preschool children and how their family support the school systems in Calgary correlated to the number of crimes, disorders and traffic incidents in a community apart from other attributes that we discussed in fig. 18. We see in the results that the crimes and disorders are not correlated to the number of preschool children, but they showed slight correlation to the number of dwellings that support public school system and both the public and separated school system. It is interesting that the crimes and disorders are strongly correlated to the number of dwellings with unknown support of school system or unable to determine. It might hint that the more the number of families that do not care about the school systems or not to answer such question in the census, the more number of crimes and disorders could be anticipated in a community. And for traffic incidents, the results matched the expectation that it would not be correlated to the number of preschool children or what school systems supported by the dwellings.
\begin{figure}
    \centering
    \includegraphics[width=1\hsize]{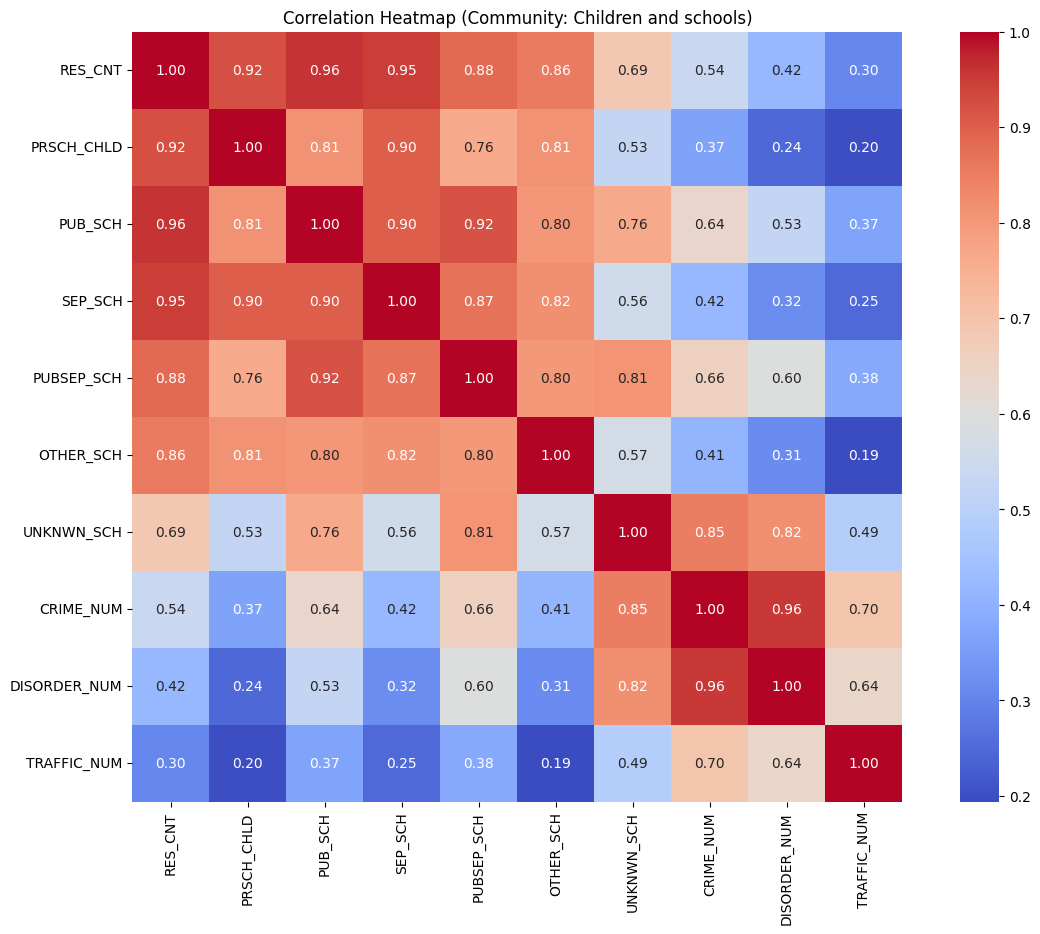}
    \caption{Correlation Heatmap (Community Children and Schools)}
    \label{fig:corr_children}
\end{figure}

For fig. 25, it shows the correlation between crimes, disorders and traffic incidents with the number of street light, their total wattage and the number of trees in a community. The number of crimes, disorders and traffic incidents show weak correlation to the total wattage of street lights in a community and no correlation to the number of trees. Such weak correlation may be the results of external correlation to some other attributes, such as the number of apartments or dwellings in a community.
\begin{figure}
    \centering
    \includegraphics[width=1\hsize]{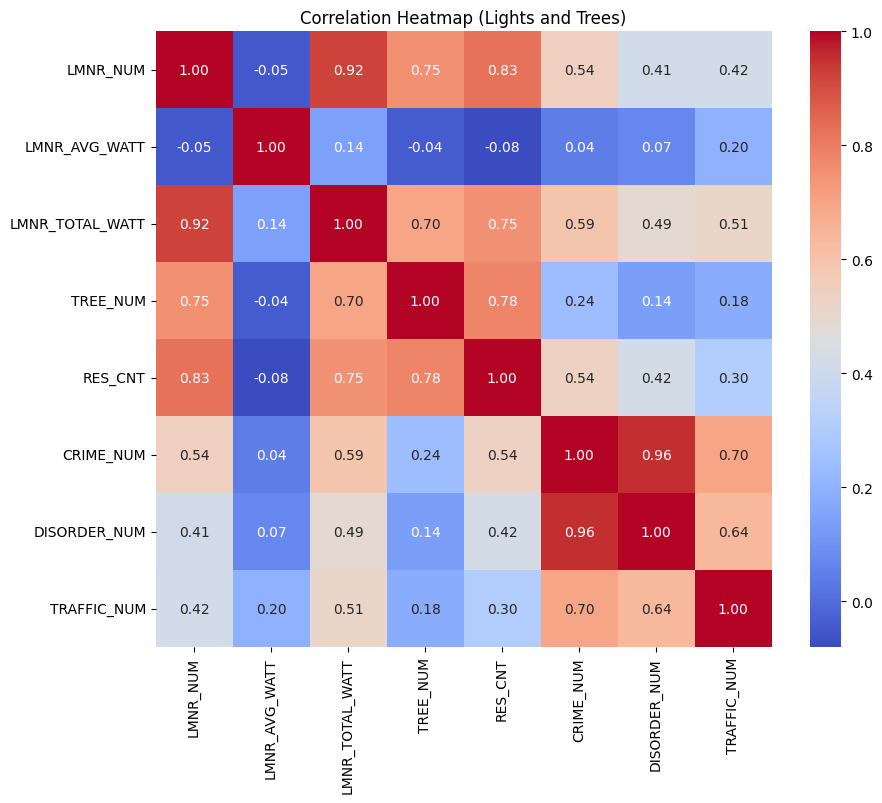}
    \caption{Correlation Heatmap (Lights and Trees)}
    \label{fig:corr_light_tree}
\end{figure}

Fig 26 shows the correlation between crimes, disorders and traffic incidents with the number of pets, cats and dogs. The correlations between the target interests and the pet attributes are weak such that we cannot derive any insights from the results. It is expected that the pet attributes in a community should not be correlated to crimes, disorder and traffic incidents, but we just want to see if the results are not as expected.
\begin{figure}
    \centering
    \includegraphics[width=1\hsize]{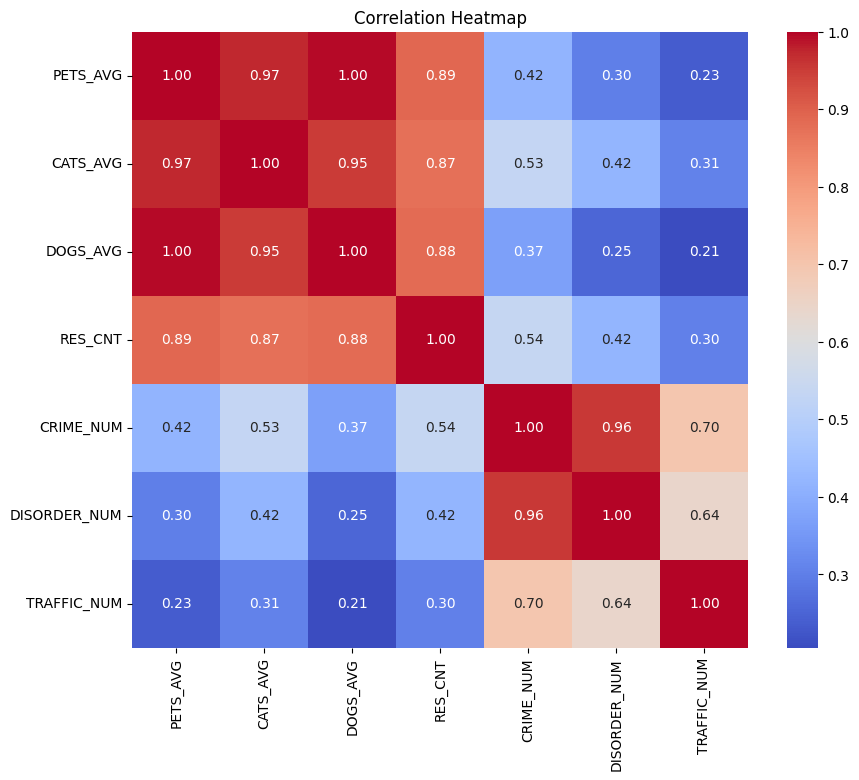}
    \caption{Correlation Heatmap (Pets)}
    \label{fig:corr_pet}
\end{figure}

\subsection{Regression Results}
For the regressions, we were not trying to build any accurate models. Instead, we leveraged the linear and random forest regressors to determine the feature importance of crimes, disorders, and traffic incidents. We fed the regressors with strongly correlated attributes from previous correlation results. For crimes and disorders, the regressors considered the number of dwellings with unknown school support dominates the regression, which are 51\% and 49\% respectively, followed by the number of persons living in an apartment, which are 23\% and 28\% respectively. And for traffic incidents, the number of crimes (37\%) and disorders (25\%) in a community dominates the regression than other attributes.

\subsection{Clustering Results}
It should be noted that clustering algorithms were only ran on the datasets that contained latitude and longitude coordinates when acquired. For the datasets where coordinates had to be calculated, clustering was not done as no meaningful results could be acquired, and clusters were non-existent. 
\subsubsection{K-Means}
We ran the K-Means clustering algorithm on six of the datasets used. The two datasets that did not receive clustering was the census data, and the community disorder data. Clustering was ran on the crime data for K-means, but it was quickly realized that no meaningful clusters would be derived since the longitude and latitude was calculated based on the community centroids.
\begin{figure}
    \centering
    \includegraphics[width=.7\hsize]{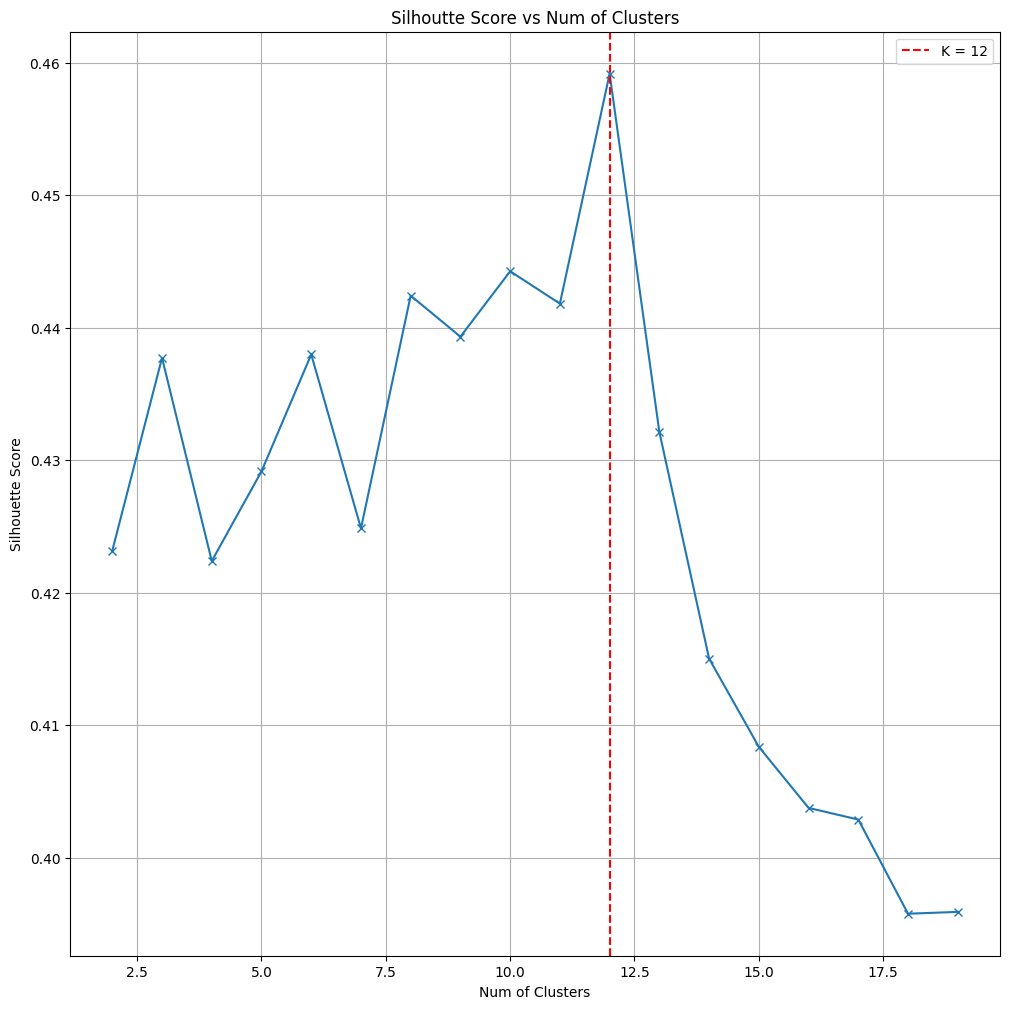}
    \caption{Silhouette Score Graph}
    \label{fig:kemeans_sil}
\end{figure}
To determine the optimal number of cluster for each dataset, silhouette scores were calculated and the cluster number with that score was assigned when the algorithm was re-ran with the optimal parameters. 
\begin{figure}
    \centering
    \includegraphics[width=0.5\hsize]{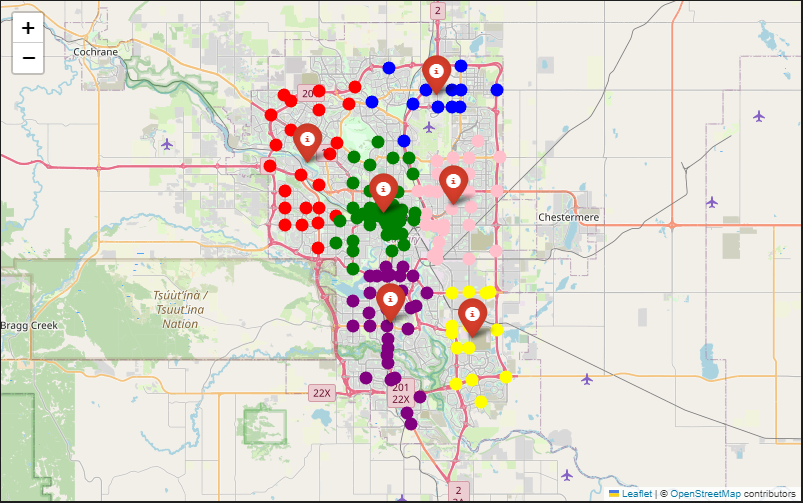}\hfill
    \includegraphics[width=0.5\hsize]{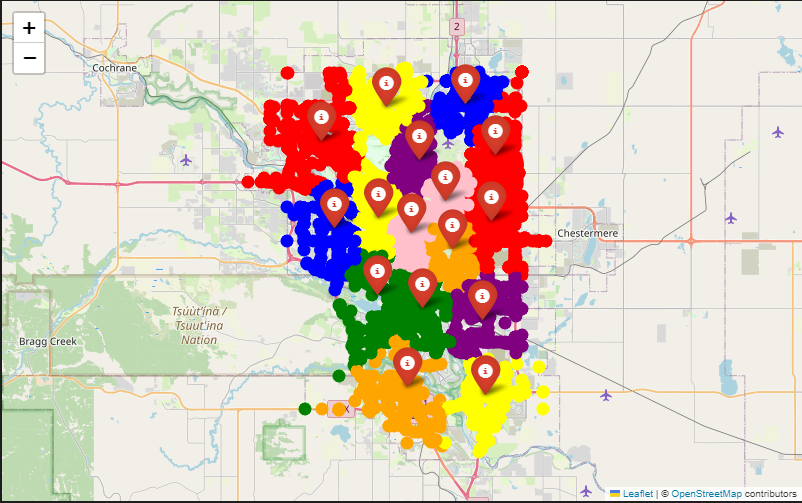}\hfill
    \\[\smallskipamount]
    \includegraphics[width=0.5\hsize]{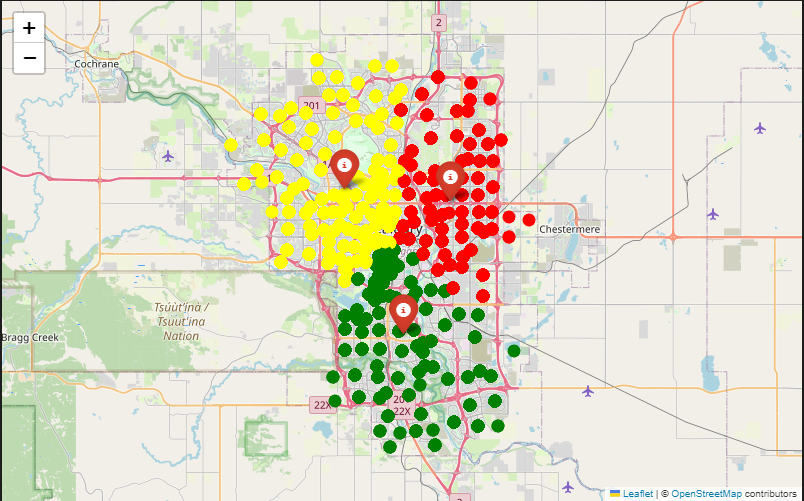}\hfill
    \includegraphics[width=0.5\hsize]{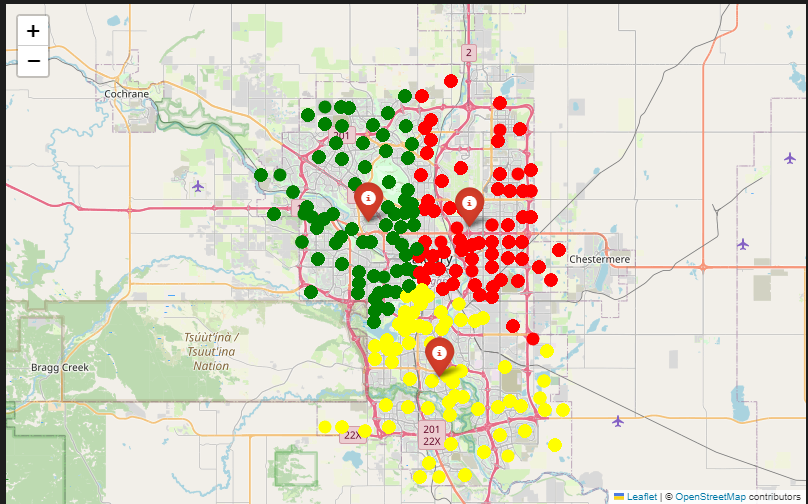}\hfill
    \\[\smallskipamount]
    \includegraphics[width=0.5\hsize]{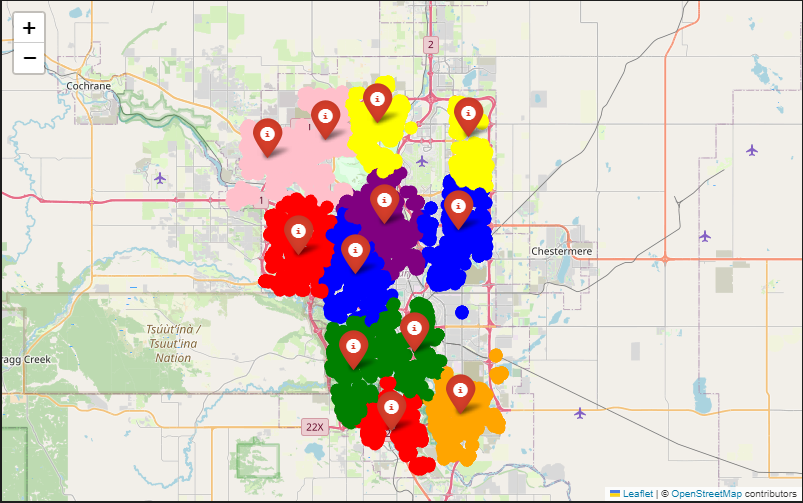}\hfill
    \includegraphics[width=0.5\hsize]{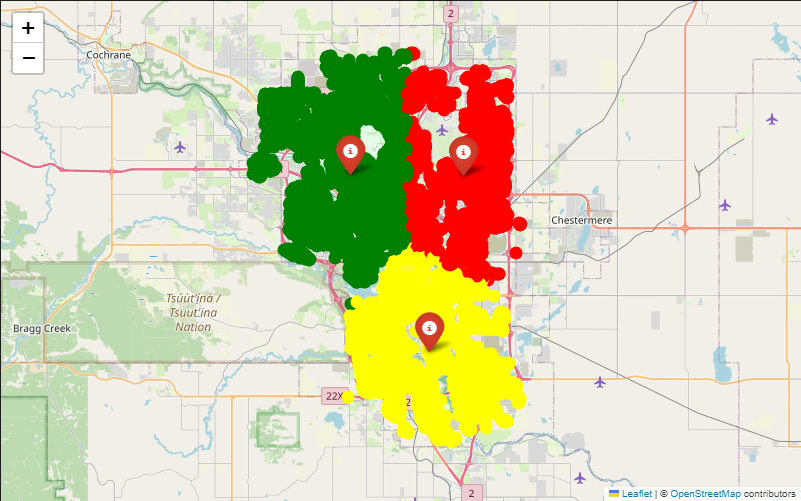}\hfill
    \\[\smallskipamount]
    \caption{K-Means Clustering Results. Top row: Traffic Camera, Traffic Incident; Middle row: Pets, Crime; Last row: Trees, Lumens}
    \label{kmeans_images}
\end{figure}
\begin{table}[htbp]
    \centering
    \begin{tabular}{|c|c|c|}
    \hline
    DataFrame & Max Silhouette Score & Number of Clusters \\
    \hline
    Lumens & 0.4218 & 3 \\
    Trees & 0.4591 & 12 \\
    Traffic Incidents & 0.3802 & 16 \\
    Crime & 0.3941 & 3 \\
    Pets & 0.3813 & 3 \\
    Traffic Cameras & 0.4398 & 6 \\
    \hline
    \end{tabular}
    \caption{K-Means Results}
    \label{tab:km_tuning_results}
\end{table}
Due to time constraints, the only hyper-parameter that was adjusted for K-Means was the number of clusters, the other parameters were assumed to be 5 iterations, a random state of 0 and then the algorithm was only ran on one centroid seed. These decisions were made to speed up computation time and because the data showcased poor clustering results during the initial setup phase. 

\subsubsection{CLARANS Clustering }
After the K-Means results were computer, some small changes were made to the way clustering was done in order to reduce computational time. The major change was that the crime dataset and traffic camera dataset would no longer have clustering algorithms ran on them. The reason for this was discussed above, but to re-iterate is that no meaningful clusters would be derived from these datasets due to the nature of the latitude and longitude values within them. 
\begin{figure}
    \centering
    \includegraphics[width=0.5\hsize]{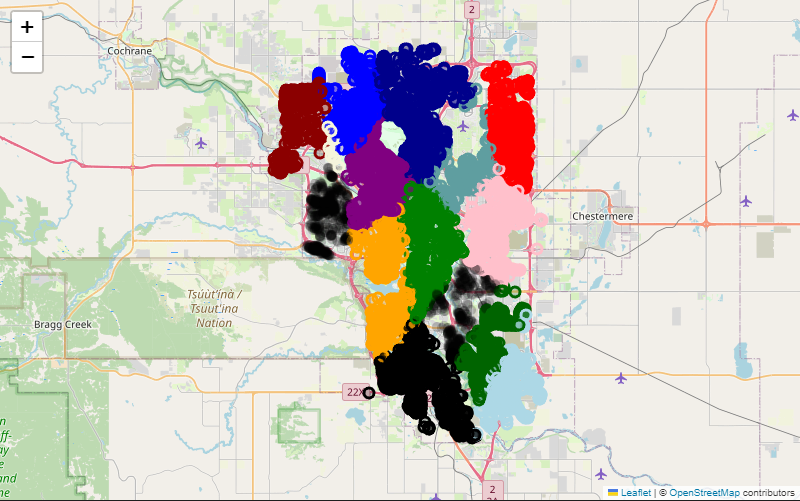}\hfill
    \includegraphics[width=0.5\hsize]{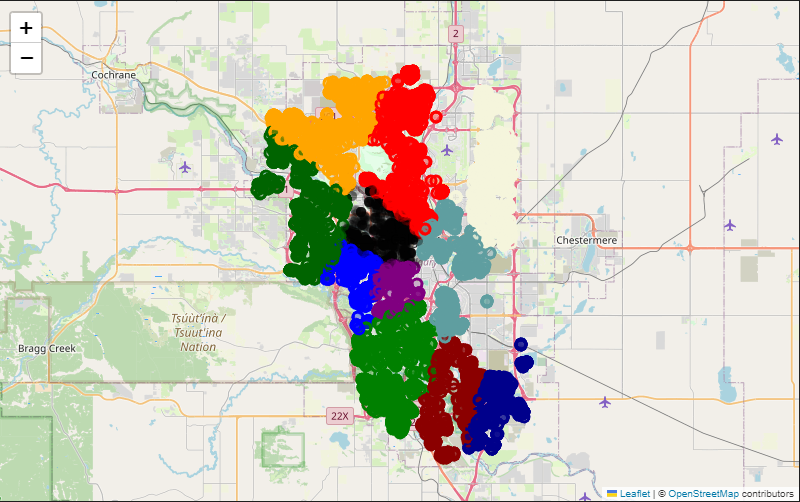}\hfill
    \\[\smallskipamount]
    \includegraphics[width=0.5\hsize]{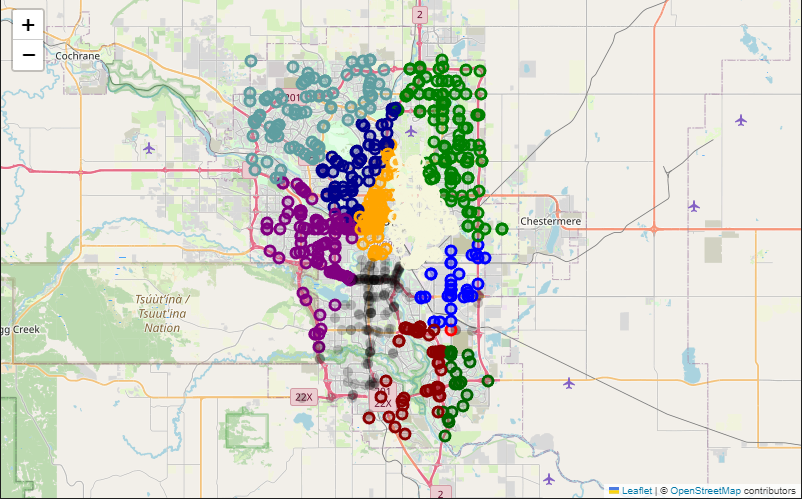}\hfill
    \includegraphics[width=0.5\hsize]{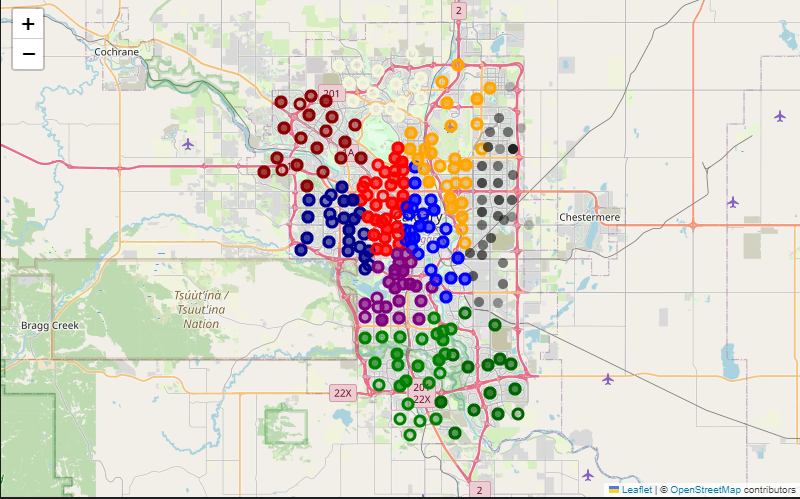}\hfill
    \caption{CLARANS Clustering Results. Clockwise from top-right: Lumens, Trees, Pets, Traffic Incidents}
    \label{fig:clarans-result}
\end{figure}
The results of the CLARANS clustering algorithm are of a higher quality than K-Means, but still not great overall. The only dataset that has what would be considered good clustering results is the dataframe containing all of the public trees in the City of Calgary. This can be seen in the top right map of Figure 30, where there are some distinct clusters to be found in the map. Otherwise, none of the dataframes have a high silhouette score indicating that there is not much clustering to be found in the datasets - alike K-Means.
\begin{table}
    \centering
    \footnotesize 
    \setlength{\tabcolsep}{2pt} 
    \begin{tabular}{|c|c|c|c|c|}
    \hline
    DataFrame & Silhouette Score & Clusters & Local Minima & Neighbours \\
    \hline
    Lumens & 0.5552 & 15&6&6 \\
    Trees & 0.8097 & 11&9&3 \\
    Traffic Incidents & 0.4027 & 12&10&9 \\
    Pets & 0.3886 & 10&9&5 \\
    \hline
    \end{tabular}
    \caption{CLARANS Results}
    \label{tab:param_tuning_results}
\end{table}

\subsubsection{DBSCAN Clustering}
DBSCAN clustering is by far the worst algorithm that is implemented on the datasets, this is because for all intents and purposes no matter the dataset, there is only one cluster to be found. This indicates that all of the data is in close spatial proximity to each other, which is this true, but it also means that it is impossible to make any insights from clustering through DBSCAN. 
\begin{figure}
    \centering
    \includegraphics[width=0.5\hsize]{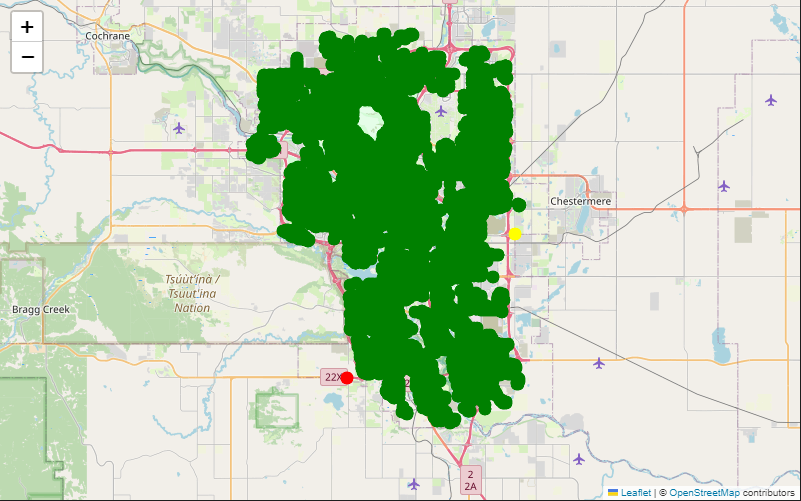}\hfill
    \includegraphics[width=0.5\hsize]{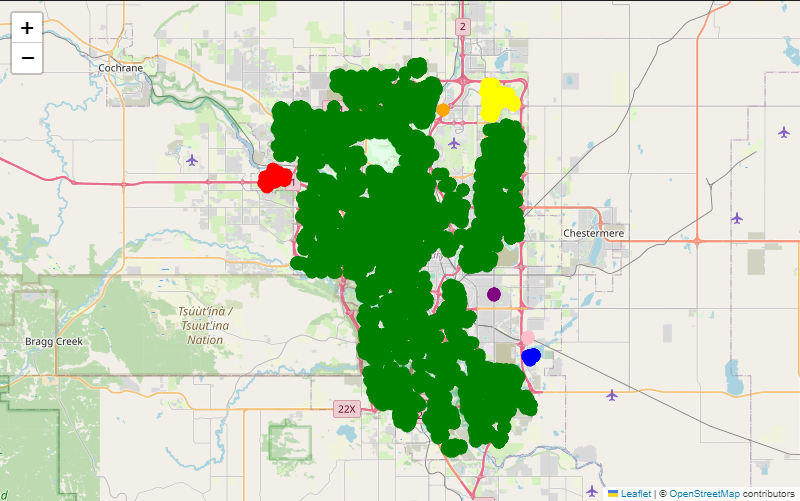}\hfill
    \\[\smallskipamount]
    \includegraphics[width=0.5\hsize]{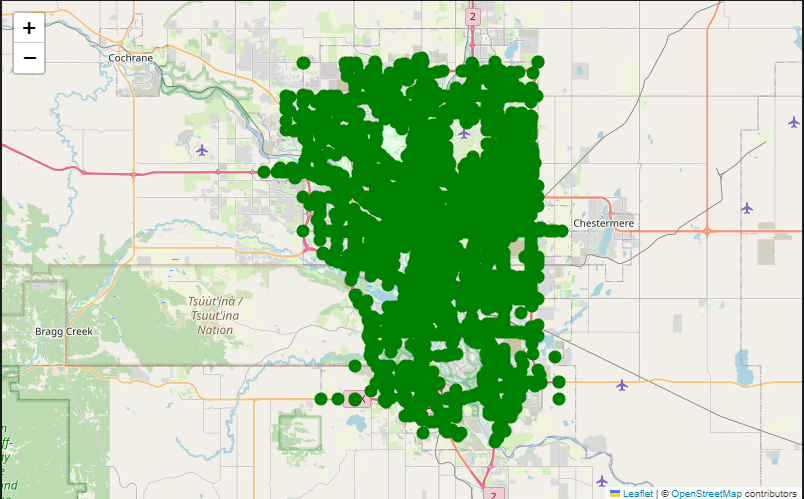}\hfill
    \includegraphics[width=0.5\hsize]{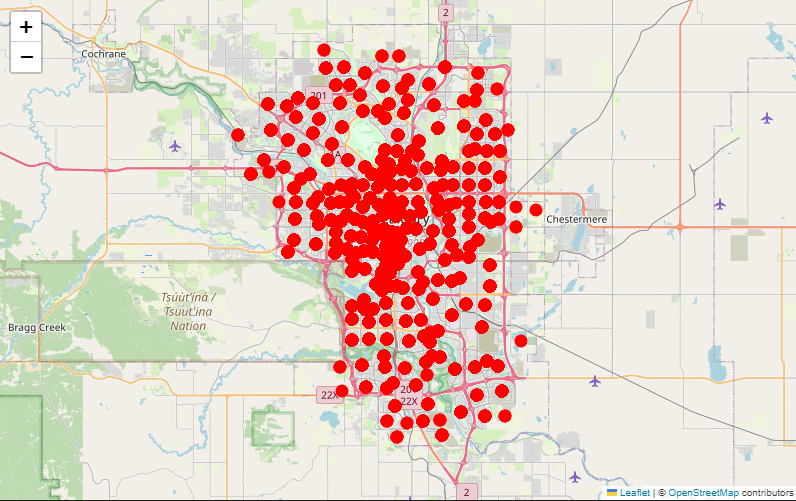}\hfill
    \\[\smallskipamount]
    \includegraphics[width=0.5\hsize]{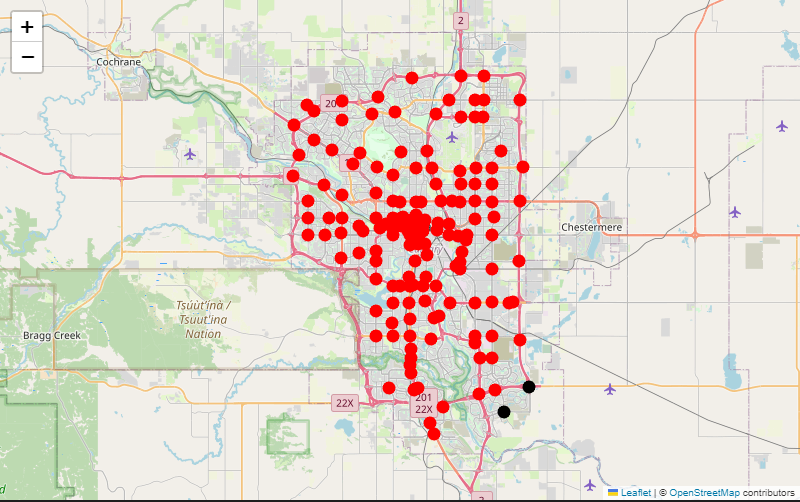}\hfill
    \caption{DBSCAN Clustering Results. Top row: Lumens, Trees; Middle row: Traffic incidents, Pets; Bottom row: Traffic cameras}
    \label{fig:dbscan-clustering-result}
\end{figure}

Looking at Table 3, the Pets dataframe has a very high silhouette score, but the EPS is so low it makes the entire dataframe into one cluster, so the high silhouette score is more or less negated by the fact that no clustering analysis can be done. 
\begin{table}
    \centering
    \begin{tabular}{|c|c|c|c|}
    \hline
    DataFrame & Max Silhouette Score & eps & min\_samples \\
    \hline
    Lumens & 0.4161 & 0.006 & 2 \\
    Trees & 0.5680 & 0.009 & 2 \\
    Traffic Incidents & 0.3739 & 0.030 & 2 \\
    Pets & 0.9713 & 0.0001 & 2 \\
    Traffic Cameras & 0.3815 & 0.070 & 18 \\
    \hline
    \end{tabular}
    \caption{DBSCAN Results}
    \label{tab:dbscan_results}
\end{table}

\subsubsection{Hierarchical Clustering}
Much alike the other algorithms, hierarchical clustering does not produce very high silhouette scores, nor very good results. Granted, Hierarchical clustering at least produces distinct clusters for all the tested dataframes, but not much meaningful information can be extracted from them. Once again further proving that clustering, no matter the method does not work well on the selected datasets. 
\begin{figure}
    \centering
    \includegraphics[width=0.5\hsize]{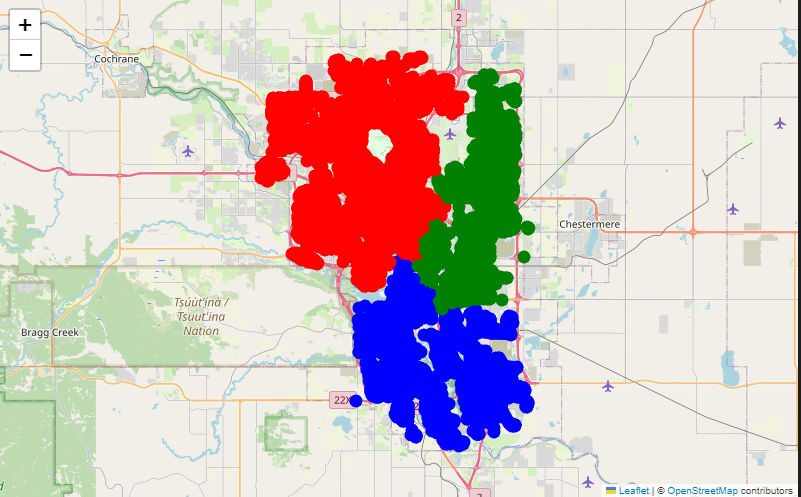}\hfill
    \includegraphics[width=0.5\hsize]{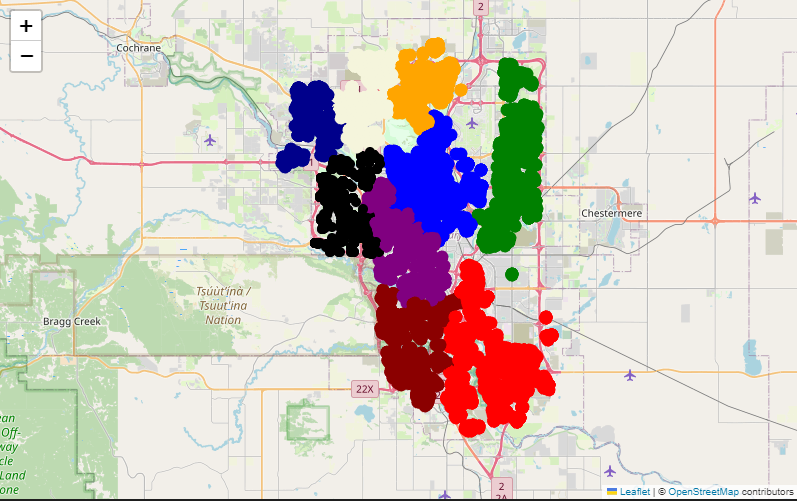}\hfill
    \\[\smallskipamount]
    \includegraphics[width=0.5\hsize]{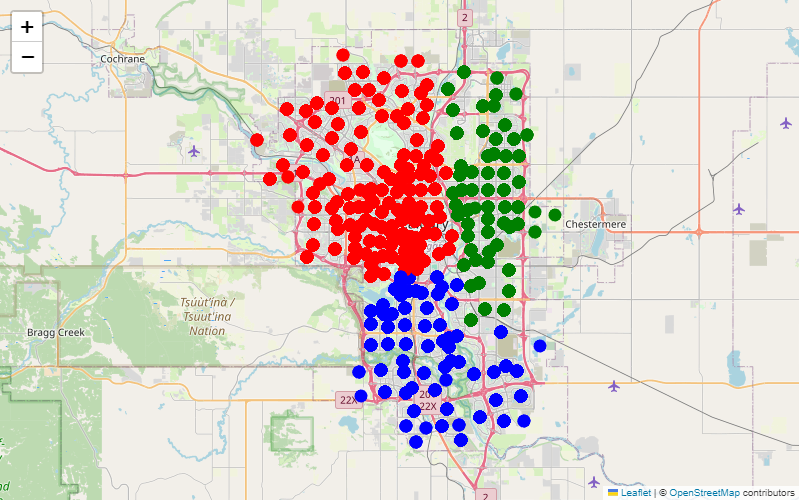}\hfill
    \includegraphics[width=0.5\hsize]{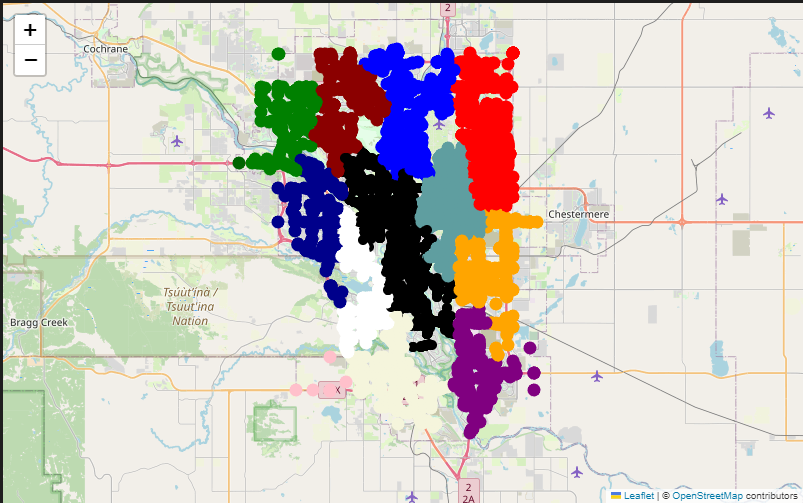}\hfill
    \caption{Hierarchical Clustering Results. Clockwise from top left: Lumens, Trees, Traffic Incidents, Pets}
    \label{fig:hierarchical-clustering-results}
\end{figure}
\begin{table}[htbp]
    \centering
    \begin{tabular}{|c|c|c|c|}
    \hline
    DataFrame & Linkage & Number of Clusters & Silhouette Score \\
    \hline
    Lumens & ward & 3 & 0.4500 \\
    Trees & ward& 9 & 0.4379 \\
    Traffic Incidents & average & 14 & 0.3690 \\
    Pets & complete & 3 & 0.3721 \\
    \hline
    \end{tabular}
    \caption{Hierarchical Clustering Results}
    \label{tab:hclustering_results}
\end{table}

\subsubsection{CLIQUE Clustering}
\begin{figure}
    \centering
    \includegraphics[width=0.5\hsize]{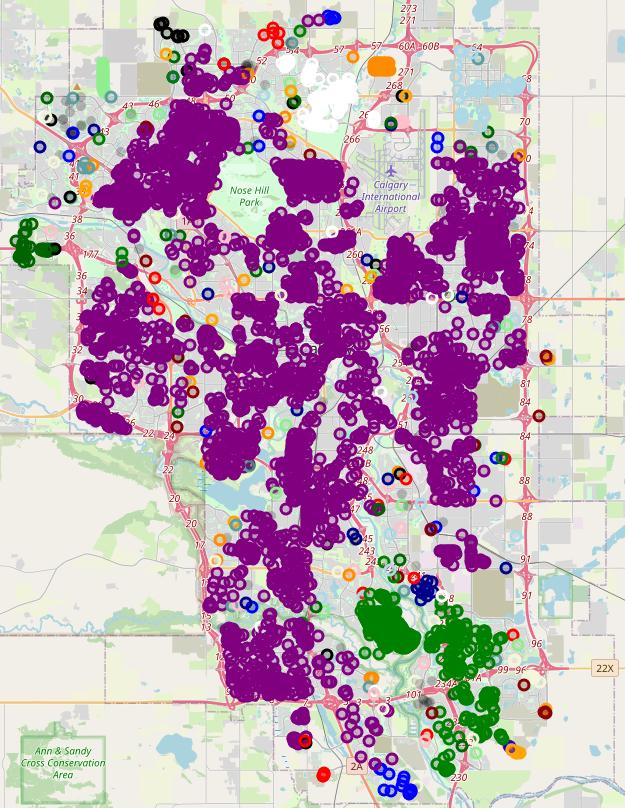}\hfill
    \includegraphics[width=0.5\hsize]{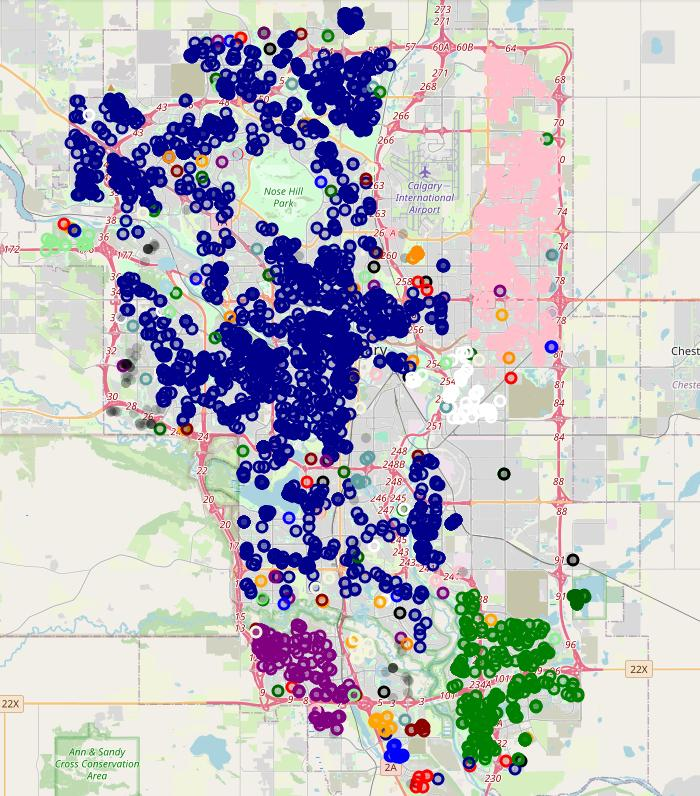}\hfill
    \\[\smallskipamount]
    \includegraphics[width=0.5\hsize]{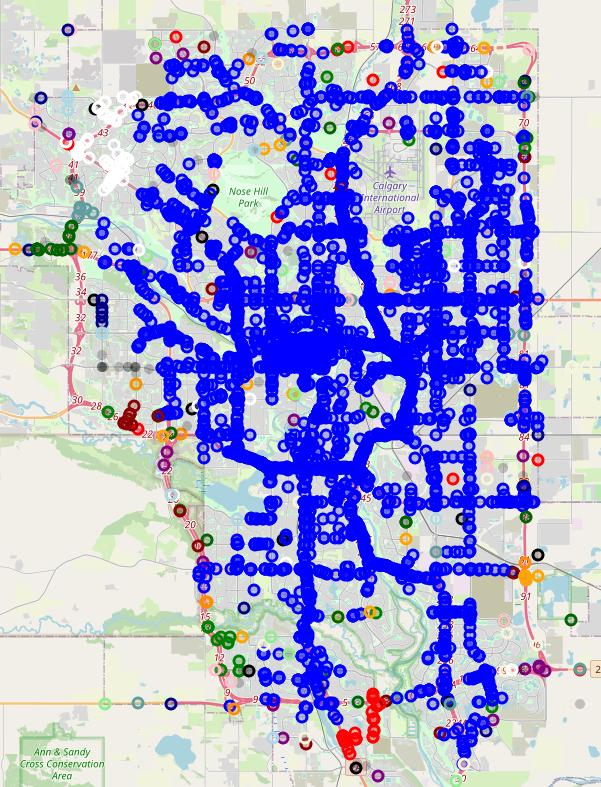}\hfill
    \includegraphics[width=0.5\hsize]{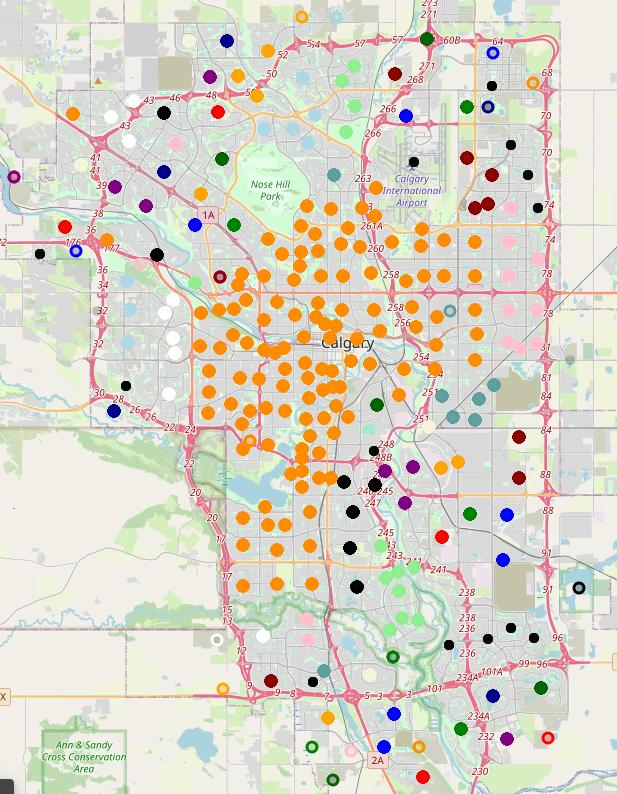}\hfill
    \caption{Clique Clustering Results. Clockwise from top left: Lumens, Trees, Pets, Traffic Incidents}
    \label{fig:clique-label}
\end{figure}
\begin{table}[htbp]
    \centering
    \begin{tabular}{|c|c|c|c|}
    \hline
    DataFrame & Silhouette Score & Interval & Threshold \\
    \hline
    Lumens & 0.1220 & 18 &0  \\
    Trees & 0.1220 & 18 &0 \\
    Traffic Incidents & 0.4045  & 14& 0 \\
    Pets &  0.3448 & 10& 0 \\
    \hline
    \end{tabular}
    \caption{Clique Clustering Results}
    \label{tab:clique_results}
\end{table}
Overall, Clique clustering does not perform well in clustering the data. It can be seen in all the images that one of the clusters dominates the majority of the points in the dataframe. This makes deriving any results from the figure extremely challenging as the Clique algorithm is indicating that there is one major cluster in each dataset with a couple smaller peripheral clusters. This is quite similar to DBSCAN, which also produced very poor results.
\subsubsection{OPTICS Clustering}
\begin{figure}
    \centering
    \includegraphics[width=0.5\hsize]{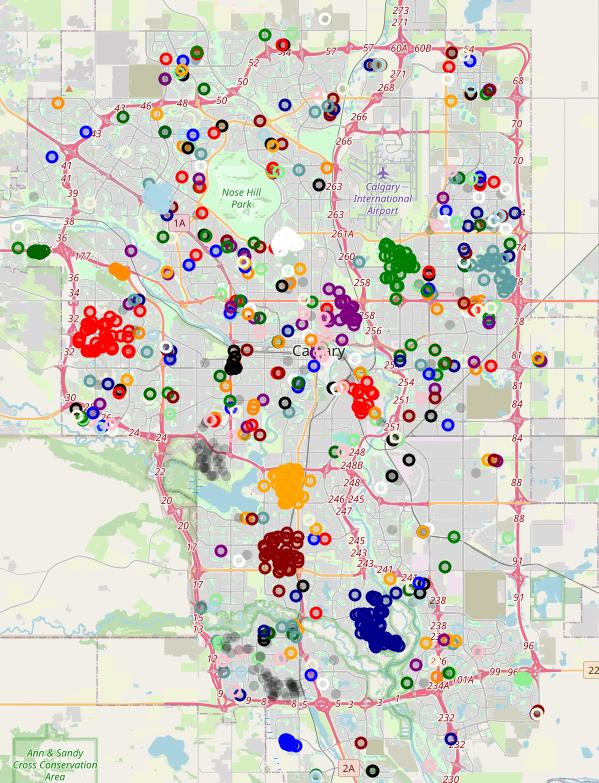}\hfill
    \includegraphics[width=0.5\hsize]{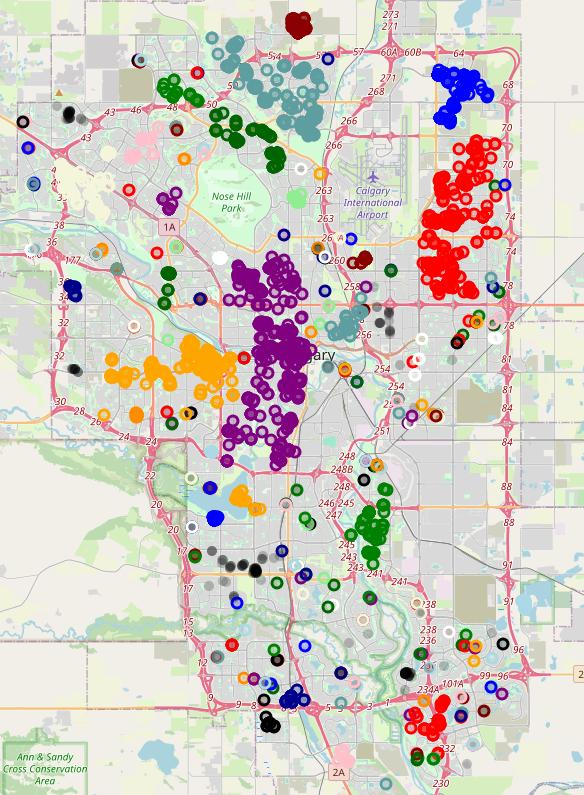}\hfill
    \\[\smallskipamount]
    \includegraphics[width=0.5\hsize]{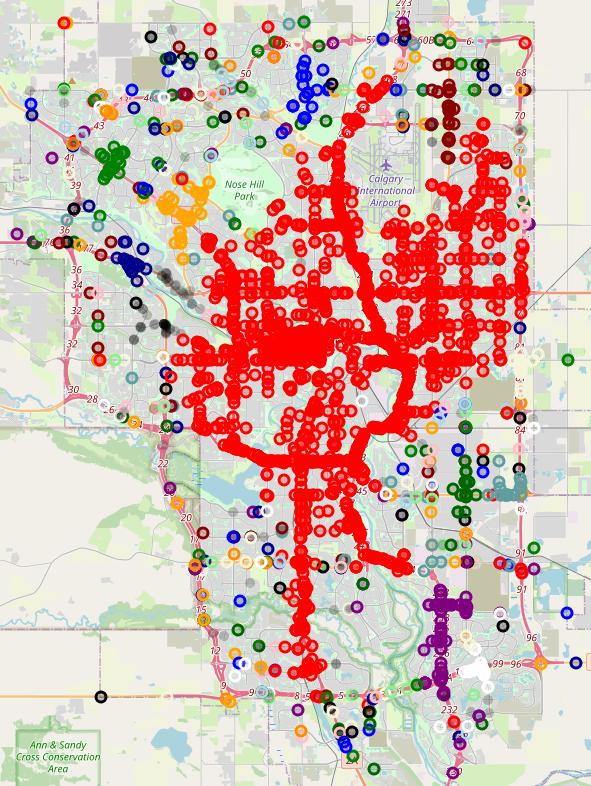}\hfill
    \includegraphics[width=0.5\hsize]{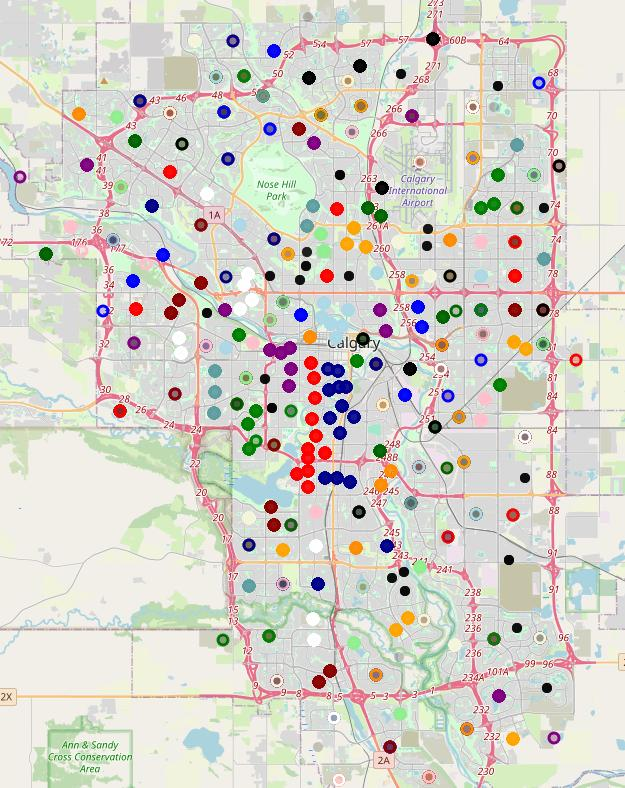}\hfill
    \caption{Optics Clustering Results. Clockwise from top left: Trees, Lumens, Pets, Traffic Incidents}
    \label{fig:Optics-label}
\end{figure}
\begin{table}[htbp]
    \centering
    \footnotesize 
    \setlength{\tabcolsep}{3pt} 
    \begin{tabular}{|c|c|c|c|}
    \hline
    DataFrame & Silhouette Score & Radius & Neighbours \\
    \hline
    Lumens & -0.0990 & 0.01 & 8 \\
    Trees & 0.1995 & 0.01 & 8 \\
    Traffic Incidents & -0.3812 & 0.01 & 8 \\
    Pets & 0.3095 & 0.01 & 8 \\
    \hline
    \end{tabular}
    \caption{Optics Clustering Results}
    \label{tab:Optics_results}
\end{table}

Optics clustering produces extremely confusing results, with the Lumens dataset, it suggests that there may be clusters of streetlights throughout the city. This is not supported by other clustering algorithms but is interesting nonetheless. It also suggests that there are clusters of trees throughout the city, which is again not overly supported by the other algorithms. Then the next two results are not sufficient for any further analysis, where the clusters are either spanning giant portions of the city or are confined to individual points. Two of the silhouette scores are also in the negative, which indicates that hyper-parameters that are being used and tested. Due to time constraints, there was not time to test further hyper-parameters to confirm this theory. 

\section{Discussion and Future Work}
\subsection{Key Findings}
This study utilized data analysis and data mining techniques to delve into the crucial factors influencing community safety in the city of Calgary. 

Through the exploratory data analysis, we revealed patterns, insights and trends in geospatial, categorical and temporal perspectives.

While for correlation analysis, we found significant correlations between crime rates, disorder incidents, traffic accidents, and several community attributes, including population, housing types, streetlight quantity, and tree quantity. It revealed the intrinsic connections between different community attributes and the 3 main concerns. For instance, the population related attributes demonstrated positive correlations with crimes and disorders, while the apartment related attributes exhibited strong positive correlations with crimes and disorders. This implies that the housing structure may be primary factors influencing community crime. On the other hand, crimes and disorders are strongly correlated to the number of dwellings with unknown support of the school system or unable to determine. It implies that, in a community, the more families not caring about the education systems, there could be potentially more crimes and disorders.

And by regression analysis, we tried to find out which attribute contribute mostly on the crimes, disorders and traffic incidents. The results revealed that the number of families that don't care about the education system is contributing a large part in regressing the crimes and disorders. It implies that how people care about the education system in Calgary or the education system itself could be a key factor in crimes and disorders of the city. For traffic incidents, it is as expected the same as the results of correlation analysis, which is depending on the crimes and disorders instead of attributes in a community.

We have also tried on clustering data mining technique. Overall, clustering results were very poor across the board. K-Means overall produced the best results out of all the algorithms, but even then the highest silhouette score was only 0.4, which is not an optimal result. As all clustering algorithms do a poor job at producing distinct cluster across all the datasets, the overall indication is that the data is not overtly clustered or the hyper-parameters are not correctly optimized. OPTICS algorithm showed an interesting characteristic where the silhouette score stayed constant over different set of hyperparameters; which can suggest that data is very difficult to cluster, or there is an insufficient variation in hyperparameter, or our algorithm has reached a local optimization plane and huge shift in the hyperparameter is needed to get out of it.
Most of the clustering gave us a negative silhouette score. This happens when the mean distance between a sample and all other points in the same cluster is greater than the mean distance between that sample and all other points in the next nearest cluster, meaning that the clustering algorithm has assigned a data point to the wrong cluster, as it is more similar to points in another cluster than to points in its own cluster. This further indicates that the clusters are overlapping or poorly separated.

The data cleaning process also has a flaw when it comes to producing clustering results. For the datasets, such as crime, that did not have a latitude and longitude originating from the dataset, are unable to produce meaningful clustering. This is because the same community centroid coordinates were provided for every entry that was located in that community. Causing there to be not enough data points for there to be any meaningful clustering results. This essentially rendered three of the seven datasets for Calgary to be unusable for clustering algorithms. Another issue that arose is that from the selected parameters is that a few number of clusters were found to be optimal, making it incredibly challenging to find any significant clusters when each cluster is covering a large portion of the city. 

\subsection{Innovations}
The innovations of this study lie in: (1) integrating multiple heterogeneous data sources including streetlights, trees, crime, and traffic to depict community safety from multidimensional perspectives; (2) introducing advanced clustering algorithms and machine learning models to reveal the distribution patterns and influencing mechanisms of safety issues from both spatial and attribute perspectives; (3) applying techniques such as feature importance analysis to interpret the correlation analysis, providing actionable intelligence for community safety governance.

\subsection{Limitations and Future Directions}
Due to time and data availability constraints, this study only covered limited community indicators and data within a short-term timeframe. Future research could collect more comprehensive data and conduct long-term spatio-temporal and Trajectory analyses. Additionally, the causes of urban safety issues are complex and diverse, making it difficult to elucidate the causal relationships and mechanisms between various factors solely from a data perspective. Subsequent research could integrate qualitative methods such as expert interviews and case studies to explore the socioeconomic, institutional, and cultural factors influencing community safety.

While our study focused on Calgary, the analysis framework and technical roadmap can be generalized to other cities. Conducting comparative studies across multiple cities would help discover common patterns, facilitate knowledge exchange, and contribute wisdom to building safer and more inclusive urban environments.

\section{Conclusion }
In conclusion, this study utilized data mining techniques to analyze key factors affecting community safety in Calgary from multiple perspectives, revealing spatial disparities in safety levels and establishing effective predictive models. These findings not only fill the gap in quantitative urban research but also provide new scientific evidence to urban managers and policymakers, supporting fine-grained community safety governance. As we move towards "smart cities," data mining undoubtedly will play a crucial role. We look forward to more researchers engaging in this field, leveraging the power of data to guide the construction of safer cities.

\end{document}